 \documentclass[final,5p,twocolumn,authoryear]{elsarticle}
 \nopreprintlinetrue

\usepackage{xcolor}
\usepackage{enumerate}
\usepackage{etoolbox}
\usepackage{mathtools}
\usepackage{amssymb}
\usepackage{amsmath}
\usepackage{multirow}
\usepackage{booktabs}
\usepackage{amssymb}
\usepackage[hidelinks]{hyperref}
\usepackage{flushend}
\usepackage{amsfonts}
\usepackage{times}
\usepackage{makecell}
\usepackage{amsthm}
\usepackage[framemethod=TikZ]{mdframed}
\usepackage{caption}

\newtheorem{problem}{Problem}[]

\usepackage{tikz}
\usetikzlibrary{shapes,arrows}
\usetikzlibrary{arrows,calc,positioning}

\tikzset{
	block/.style = {draw, rectangle,
		minimum height=1.2cm,
		minimum width=1.2cm},
	input/.style = {coordinate,node distance=1cm},
	output/.style = {coordinate,node distance=2cm},
	arrow/.style={draw, -latex,node distance=2cm},
	pinstyle/.style = {pin edge={latex-, black,node distance=1cm}},
	sum/.style = {draw, circle, node distance=1cm},
}

\definecolor{backgreen}{HTML}{E9F3DF}
\definecolor{backblue}{HTML}{DAE5EC}
\definecolor{backpurple}{HTML}{E5E0E8}
\definecolor{backorange}{HTML}{F2E9DF}

\mdfdefinestyle{callout}{%
	linecolor=black,
	linewidth=0pt,%
	roundcorner=4pt,
	innertopmargin=7pt,
	innerbottommargin=7pt,
	innerrightmargin=7pt,
	innerleftmargin=7pt,
	leftmargin = 0pt,
	rightmargin = 0pt,
	backgroundcolor=backorange
}

\mdfdefinestyle{example}{%
	frametitleaboveskip = 8pt,
	frametitlebelowskip = 2pt,
	frametitlefont = \normalfont\textit,
	linecolor=black,
	linewidth=0pt,%
	roundcorner=4pt,
	innertopmargin=4pt,
	innerbottommargin=7pt,
	innerrightmargin=7pt,
	innerleftmargin=7pt,
	leftmargin = 0pt,
	rightmargin = 0pt,
	backgroundcolor=backgreen
}

\mdfdefinestyle{prospects}{%
	frametitleaboveskip = 8pt,
	frametitlebelowskip = 2pt,
	frametitlefont = \normalfont\textit,
	linecolor=black,
	linewidth=0pt,%
	roundcorner=4pt,
	innertopmargin=4pt,
	innerbottommargin=7pt,
	innerrightmargin=7pt,
	innerleftmargin=7pt,
	leftmargin = 0pt,
	rightmargin = 0pt,
	backgroundcolor=backblue
}

\AtBeginEnvironment{quote}{\par \it}

\newcommand{\rev}[1]{{\leavevmode\color{black}#1}}
\newcommand{\revM}[1]{{\leavevmode\color{black}#1}}
\newcommand{\revm}[1]{{\leavevmode\color{black}#1}}
\newcommand{\revp}[1]{{\leavevmode\color{black}#1}}
\newcommand{\revd}[1]{{\leavevmode\color{black}#1}}

\newcommand{\rmr}[1]{{\leavevmode\color{black}#1}}

\newif\ifmargincomments 
\margincommentstrue

\ifmargincomments

\else

\fi

\begin{document}
\begin{frontmatter}
\title{\rev{Distributed Design \revM{of} \revm{Ultra} Large-Scale Control} \revm{Systems}:\\Progress, Challenges, and Prospects}
\author{Leonardo~Pedroso\fnref{addressTUe,addressISR}\corref{correspondingAuthor}} 
\ead{l.pedroso@tue.nl}
\author{Pedro~Batista\fnref{addressISR}}
\author{W.P.M.H.~(Maurice)~Heemels\fnref{addressTUe}}
\address[addressTUe]{Control Systems Technology section, Eindhoven University of Technology, The Netherlands}
\address[addressISR]{Institute for Systems and Robotics, Instituto Superior T\'ecnico, Universidade de Lisboa, Portugal\vspace{-0.4cm}}
\cortext[correspondingAuthor]{Corresponding author.}

\begin{abstract}
The transition from large centralized complex control systems to \rev{distributed} configurations that rely on a network of a very large number of interconnected simpler \revm{subsystems} is ongoing and inevitable in many applications. It is attributed to the quest for resilience, flexibility, and scalability in a multitude of engineering fields with far-reaching societal impact. Although many design methods for \rev{distributed and decentralized control} systems are available, most of them rely on a \rev{centralized} \rev{design} procedure requiring some \revm{form} of global information of the whole system. Clearly, beyond a certain scale of the network, these centralized \rev{design procedures for distributed} controllers are no longer feasible and \revm{we refer to the corresponding systems as \textit{ultra}} \textit{large-scale systems} \revm{(ULSS)}. \revm{For these ULSS, design algorithms are needed} that are distributed \revm{themselves} among the \revM{subsystems} and \revM{are} subject to stringent requirements regarding communication, computation, and memory \rev{usage} \revM{of} each \revM{subsystem}. In this paper, a set of requirements is provided that assures a feasible real-time implementation \rev{of \revM{all phases of }a control solution} on \revm{an ultra} \rev{large scale}. State-of-the-art approaches are reviewed in \revm{the} light of these requirements and the challenges hampering the development of befitting control algorithms are pinpointed. Comparing the challenges with the current progress leads to the identification and motivation of promising research directions.


	
\end{abstract}


\end{frontmatter}

\section{Introduction}\label{sec:introduction}

Over the past several decades, the fast-paced development of new sensors, robot technologies, and control algorithms has enabled the automation of numerous \revm{tasks}. For instance, mapping the magnetic field of the Earth, analyzing the soil composition of Mars, and manufacturing mechanical parts with micrometer precision would be very challenging without the use of robotic systems equipped with cutting-edge sensors, actuators, and control algorithms. Although in many settings simple  \rev{centralized} architectures are \revm{still} used, there is a trend towards more decentralized and networked control solutions and systems-of-systems configurations with growing numbers of subsystems in many application domains. Consider, as an example of this trend, the task of providing global connectivity services, which has historically been carried out resorting to constellations of less than one hundred satellites in medium or geosynchronous Earth orbit. 
\begin{figure}[t]
	\centering
	\includegraphics[width = 0.65\linewidth]{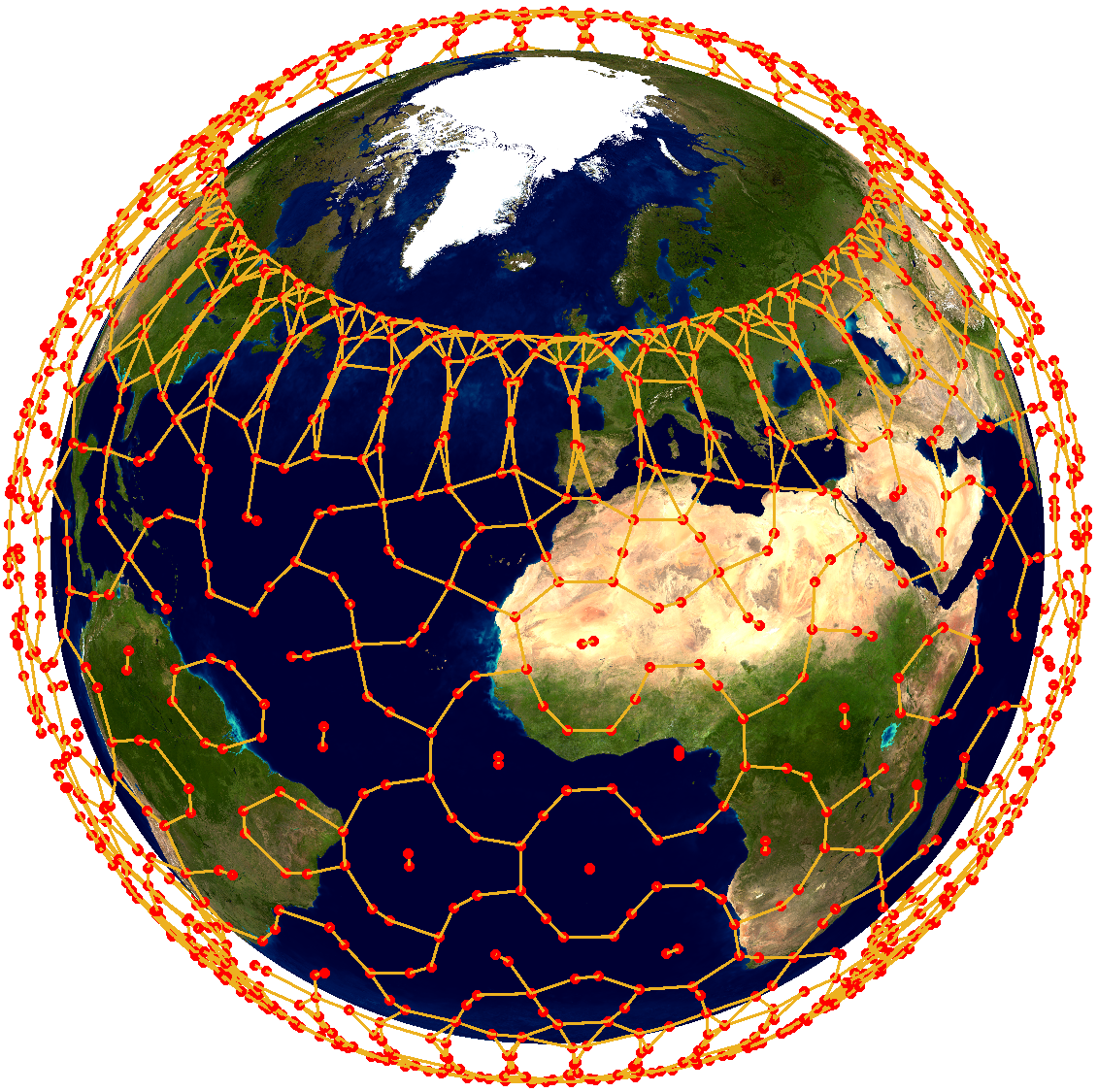}
	\caption{Satellites in one of the shells of the Starlink mega-constellation with communication links depicted as yellow edges.}
	\label{fig:starlink}
	\vspace{-0.5cm}
\end{figure}
\revm{The} soaring demand for low-latency high-capacity reliable global broadband connectivity motivated the recent transition towards the development of  \revm{mega-constellations} comprised of tens of thousands of satellites in low Earth orbit. \revm{The trend towards} scalability and reliability is common across several applications and has initiated the transition towards architectures of \revm{an ultra large} number of systems that coordinate locally to achieve a global goal. \rev{In the following three subsections, we describe the systems that fall within this framework, we zoom into an emerging subclass that we call \revm{ultra} large-scale systems \revm{(ULSS)}, and we outline the goal and scope of this vision paper.}

\subsection{\rev{Systems-of-systems}}\label{sec:sos}

\rev{The systems under study in this paper are \emph{systems-of-systems} \revM{(also called multi-agent systems)} that interact in a network and coordinate to serve a common purpose. Unsurprisingly, this emerging framework arises with powerful inspiration \revm{from} nature, mimicking large numbers of animals working towards a common goal to achieve tasks that are beyond the capabilities of a single being \revM{\citep{KubeZhang1993}}. Examples are ant colonies \citep{PrattMallonEtAl2002}, which collaboratively hunt and transport prey that is much larger than what a single ant could ever handle; honey bee swarms \citep{SeeleyVisscher2004}, \revm{whereby several scouting bees look for a nest-site in parallel and reach a consensus;} and fish schools \citep{SumpterKrauseEtAl2008}, \revm{which benefit from information exchange between fish to decide on the heading of the school}. The interest and potential of the \emph{system-of-systems} architecture for engineering applications} are driven by three pivotal properties \citep{Sahin2005}: 
\begin{itemize}\itemsep0.3em
	\item \emph{Resilience}: \rev{These} architectures are intrinsically endowed with redundancy, since the failure of a component of a \revM{subsystem} can be compensated by other \revM{subsystems} at the expense of some performance. Furthermore, each individual \revM{subsystem} is \revm{simpler} and has \revm{fewer} components \revm{than a single complex system}, becoming less prone to malfunction. This results in a resilience property, which is very compelling, especially in harsh environments, where a malfunction in a framework based on a single, multi-purpose, very complex system prompts the catastrophic failure of the complete mission;
	\item \emph{Flexibility}: \rev{The} use of several \revM{subsystems} instead of a single one allows to adopt different coordination strategies and configurations in prompt response to unpredicted changes in the environment or mission objectives;
	\item \emph{Scalability}: \rev{This} framework allows for a seamless increase \revm{(decrease)} in throughput of the task, accomplished by deploying \revm{(recalling)} additional \revM{subsystems}.
\end{itemize} \hfill

\begin{mdframed}[style=example,frametitle={\rev{\textbf{Example}: \revM{Swarms of robots}}}]
	\rev{A well-known subclass of \emph{systems-of-systems} is \revM{formed by swarms of robots}. These are systems whereby a large number of dynamically decoupled autonomous robotic \revM{subsystems} collaboratively work towards a common goal \revM{\citep{ArkinBekey1997,BrambillaFerranteEtAl2013,DorigoTheraulazEtAl2021}.}} Each \revM{subsystem} is \revm{typically rather} simple, \revm{often with} limited local sensing and communication capabilities. The aforementioned mega-constellations of satellites that provide global connectivity services fall \rev{within this subclass.}  A single shell of one of the projected mega-constellations, Starlink, is depicted in Fig.~\ref{fig:starlink}. \revM{Swarms of underwater drones for ocean sensing and swarms of aerial drones for surveillance also fall within this subclass.}
\end{mdframed}

\begin{mdframed}[style=example,frametitle={\rev{\textbf{Example}: Spatially distributed interconnected processes}}]
	\rev{Another well-known subclass of \emph{systems-of-\revM{systems}} is \revM{formed by} large-scale dynamically interconnected and spatially distributed processes.} A common example is the task of controlling the distribution of \revM{electric power} from spatially distributed sources to \rev{spatially distributed sinks} under varying demands \revM{in a smart grid}. \revm{Another example are urban traffic networks, whereby junctions, that are interconnected by road links, control the flow of vehicles, cyclists, and pedestrians with the goal of, e.g., maximizing} \revm{user throughput.}
\end{mdframed} 

\rev{Systems-of-systems, \revm{as studied in this paper,} are formally defined in Section~\ref{sec:mod_setup}. Before that, meaningful applications are detailed in Section~\ref{sec:applications}.}

\subsection{\rev{From \revm{Large-} to \revm{Ultra} Large-Scale Systems}}
\rmr{The increasing complexity of engineering problems motivates the conceptualization of increasingly large collections of simple systems. However, in this section, we argue  that although} systems-of-systems can be modeled as one single large holistic system, the increasing dimension leads to severe difficulties in designing and implementing control systems. According to \citep{bakule2008} these include:
\begin{itemize}
	\itemsep0em 
	\item Dimensionality;
	\item Information structure constraints;
	\item Uncertainty;
	\item Delays.
\end{itemize}
Instead of resisting the distributed nature of the system by attempting to design a single, very complex, centralized control system, the \emph{decentralized} and \emph{distributed control} approaches arose by embracing and exploiting the structure of the problem. Landmark references include, among others, \citet{Siljak2012,lunze1992feedback,siljak2005,bakule2008,Bakule2014}. Some particular aspects of this rich field are reviewed in Section~\ref{sec:sota}.

\rev{In this paper, we argue that beyond a certain scale of the system-of-systems, which we term \emph{\revm{ultra} large scale} \revm{(ULS)}, new disruptive techniques are needed.} Although the term \emph{\revm{ultra} large-scale system} has been loosely applied before in control settings, we propose a \revM{more accurate} definition in this section. \rev{Before doing so, we clarify the distinction  between the concepts of (i)~\emph{design} and \emph{working phases} of a control solution; and between the concepts of (ii)~\rmr{\emph{centralized},} \emph{decentralized}, and \emph{distributed} \rmr{phases}.}


First, the basic control problem for systems-of-systems \rmr{relies} on (a)~a priori information about the plant, disturbances, and design requirements; and (b)~a posteriori information about the outputs and reference signals. As a result, the control \rmr{solution} can be divided into two phases:
\begin{enumerate}[(a)]\itemsep0.3em
	\item \emph{Design} phase: Design control laws given the a priori information about the plant and the design requirements;
	\item \emph{Working} phase: Compute the control inputs given \revm{the} control \revm{laws} and the a posteriori information about the output and reference signals.
\end{enumerate}

\rmr{Second, the information structure of a phase is characterized herein using the following terms:}
\begin{enumerate}[(i)]\itemsep0.3em
	\item \rmr{\emph{Centralized} characterizes a phase that is carried out in a single computational entity that has access to global information;}
	\item  \emph{Decentralized} \rmr{characterizes a phase that is carried out by different computational entities that do not exchange information with each other;}
	\item  \emph{Distributed}  \rmr{characterizes a phase that is carried out by different computational entities that may exchange information with each other.}
\end{enumerate}
\rmr{Note that} the terms decentralized and distributed are used interchangeably by some authors and may also refer to different concepts in different control fields. \rmr{In this paper, we adopt the aforementioned definition and henceforth use these terms accordingly.} \rmr{Also notice that a degenerate case of a distributed phase whereby the communication graph of the computational entities is composed of a self-loop in each node reduces to a decentralized phase. It is also important to remark that, in a decentralized or distributed phase, computational entities may be able to infer information from one another through performance and sensor outputs despite not being able to establish direct communication. This aspect will be discussed further in Section~\ref{sec:fomulation}.}


\rev{We propose the definition of \revm{an} \emph{\revm{ultra} large-scale} system as follows:}
\begin{mdframed}[style=callout]
	A system \revM{(and the control problem associated with it)} is said to be of \revm{an} \emph{\revm{ultra} large scale}, if its dimension \revM{prevents finding a solution \revm{in} the \emph{design phase} of the control problem in a computationally feasible manner in a \emph{central} computational entity.}
\end{mdframed}
To emphasize, the difference between large-scale and \revm{ultra} large-scale systems \revm{lies} in the fact that the \rev{\emph{design}} phase of the (decentralized \rev{or distributed}) control solution can no longer be carried out on one \textit{central} computational entity. \revM{Hence}, in case a decentralized \rev{or distributed} control solution can be \rev{designed} centrally in one central device, we do not speak of \revm{an ULSS}.

\begin{mdframed}[style=example,frametitle={\rev{\textbf{Example}: Reaching \revm{an ultra} large scale}}]
	\rmr{Consider a constellation of satellites like the one depicted in Fig.~\ref{fig:starlink}. If the goal is to provide global connectivity with increasingly lower latency, the orbits need to be increasingly lower, which in turn requires an increasingly large number of satellites to provide global coverage. Coarsely, constellations in medium Earth orbit require tens of satellites, whereas constellations in low Earth orbit require tens of thousands. Let $N$ be the number of satellites of the constellation and consider} a control design method such as a linear quadratic regulator (LQR) \rmr{to maintain the shape of the constellation}. The number of operations required to compute the optimal LQR controller grows, at least, with $N^3$ \citep{RameshUtkuEtAl1989}. One can readily see that as more \rmr{satellites} are deployed (and $N$ grows), at some point the computational complexity quickly grows beyond the capabilities of a central computational entity, at which point the system becomes an ULSS.
\end{mdframed}

The transition from large- to \revm{ultra} large-scale modern complex systems has begun recently and there is no turning back. Indeed, the inevitability of this paradigm shift has been identified almost two decades ago, with particular emphasis on the computational engineering field. Indeed, in \revm{a} year-long seminal study \revm{\citep{FeilerGabrielGoodenoughEtAl2006}}, the characterization and transition towards so-called \revm{ultra} large-scale systems is comprehensively analyzed. \revm{Such} systems are characterized by complex wireless networks of thousands or tens of thousands \revm{of} sensors and decision nodes. One of the key takeaways of this study is the fact that such a transition must be accompanied by a deep revolution on design, orchestration, control, and monitoring methods and techniques:
\begin{quote}
	``We need a new science to support the design of all levels of the systems that will eventually produce the ULS systems we can envision, but not implement effectively, today.'' \textnormal{\citep[p.13]{FeilerGabrielGoodenoughEtAl2006}}
\end{quote}  
Unsurprisingly, this transition is doomed unless it is supported by a \revM{paradigm} shift in \rev{decentralized and distributed} control theory: 
\begin{mdframed}[style=callout]
	\revm{Although vast research works focus on the scalability of the \emph{working phase}, the feasibility of the \emph{design phase} \revd{on} an \revm{ULS} has been disregarded for the most part. The} 
	\revm{transition towards ULSS calls for control strategies that are scalable for the two phases.}
\end{mdframed}
 In fact, the urgent need of a paradigm change from the standpoint of the \revm{design and deployment} of \revm{ULS} control systems has been recently pointed out \revM{\citep{EngellPaulenEtAl2015,DingHanWangEtAl2019,KordestaniSafaviSaif2021,ZhanEtAl2020}}.

\subsection{\rev{Scope and Goal}}

\rev{\revM{In this paper, we} identify and formulate requirements that assure a feasible implementation \revM{of \revm{the two} phases of a} control solution on \revm{an ULSS}. \revM{We review approaches to classical decentralized and distributed control problems in \revm{the} light of their ability to satisfy these additional feasibility requirements.}}

\begin{mdframed}[style=prospects,frametitle={\revM{\textbf{Remark}: Homogenization approach}}]
	\rev{Recently, attention has been given to homogenization approaches that approximate \revm{ULSS} as a continuum of \revM{subsystems} (see \cite{FornasierSolombrino2014,BonginiFornasierEtAl2017,FornasierLisiniEtAl2019,GaoCaines2020, NikitinCanudas-de-WitEtAl2021,NikitinCanudas-de-WitEtAl2022,NikitinCanudas-de-WitEtAl2023}). Crucially, \revm{the} analysis and design of the continuum is scale-free. However, the continuum is usually described by partial differential equations, for which analysis and design \revM{can be} challenging. The nature of these techniques is fundamentally different from the classical decentralized and distributed control literature. This approach is outside the scope of this paper.}
\end{mdframed}

The literature on \rev{decentralized and distributed control} for large-scale systems is vast. The goal of this paper is not to thoroughly survey this rich field, but instead to provide insights into its evolution towards the new generation of \revm{ULS} engineering systems. To be more precise, the goal of this work is threefold: 
\begin{itemize}\itemsep0em
	\item \revM{In Sections~\ref{sec:applications} and~\ref{sec:challenges}, the} \textcolor{black}{challenges that are hampering the transition towards \revM{decentralized and distributed} control algorithms befitting \revm{an ULS} are \revM{pinpointed} and analyzed resorting to a suitable abstract modeling setup. These are underpinned with examples of pressing \revm{ULS} applications.}
	\item \revM{In Section~\ref{sec:sota}, it is investigated to what extent} \textcolor{black}{the state-of-the-art in \rev{decentralized and distributed} control has already evolved \revM{towards feasibility on \revm{an ULS}}. In fact, we will see that the number of works \revM{that suit \revm{an ULS}} is surprisingly low compared to the complete literature on \rev{this field}. On top of that, \revM{insights} into whether current techniques can be leveraged to promote this paradigm shift is given.}
	\item \revM{In Section~\ref{sec:prospects}, by} \textcolor{black}{comparing the challenges with the current \revM{state-of-the-art}, promising research directions are identified and motivated.}
\end{itemize}
\section{Applications \revM{of \revm{ultra} large-scale systems}}\label{sec:applications} 

As already mentioned in the Introduction, the quest for resilience, flexibility, and scalability in pressing applications is inducing a paradigm shift towards architectures of \revm{an ultra} large number of coordinated control systems. In this section, meaningful applications \rev{of \revm{such} systems-of-systems are discussed, \revM{see also} Table~\ref{tab:app}.}

\begin{table*}[h!]
	\centering
	\caption{Pressing applications of systems-of-systems.}
	\label{tab:app}
	\vspace{0.3cm}
	\small
	\begin{tabular}{lll}
		\toprule
		Unmanned aerial vehicle swarms & Autonomous underwater vehicle swarms & Spacecraft swarms\\
		\midrule
		\makecell[l]{
			\textbullet\ \textit{Precision agriculture}:\\ \citet{Radoglou-GrammatikisEtAl2020},\\ \citet{cobbenhagen2018heterogeneous,cobbenhagen2021opportunities}; \vspace{0.1cm}\\
			\textbullet\ \textit{Fire-fighting}: \citet{GhamryKamelZhang2017}, \\ \citet{SherstjukZharikovaEtAl2018}; \vspace{0.1cm}\\
			\textbullet\ \textit{Surveillance}: \citet{AwasthiBalusamyPorkodi2019}; \vspace{0.1cm}\\
			\textbullet\ \textit{Light shows}: \citet{AngDongEtAl2018}.
		}
		 & 
		 \makecell[l]{
		 	\textbullet\ \textit{Ocean sensing}:\\ \citet{SchulzHobsonEtAl2003}, \citet{FiorelliLeonardEtAl2006},\\ \citet{MBARI2018}, \\ \citet{LeonardPaleyEtAl2010}, \citet{mertzimekis2021radioactivity}; \vspace{0.1cm}\\
		 	\textbullet\ \textit{Sustainable deep-sea mining}: \citet{Sharma2017};\vspace{0.1cm}\\
		 	\textbullet\ \textit{Seabed mapping and exploration}:\\ \citet{TsiogkasPapadimitriouEtAl2014}, \citet{VedachalamEtAl2019}.\vspace{0.1cm}\\
	 	}
		  &  
		   \makecell[l]{
		  	\textbullet\ \textit{Earth observation}: \cite{FarragOthmanEtAl2021}; \vspace{0.1cm}\\
		  	\textbullet\ \textit{Communications}: \citet{butash2021non},\\
		  	\citet{XieZhanEtAl2021}; \vspace{0.1cm}\\
	  		\textbullet\ \textit{Astronomy}: \citet{BentumVerhoevenEtAl2009},\\\citet{DekensEngelenEtAl2014}; \vspace{0.1cm}\\
	  		\textbullet\ \textit{Planetary exploration}:\\ \citet{Petrovsky2022}.
  		}\\
		\bottomrule
	\end{tabular}\\
	\vspace{0.5cm}
	\begin{tabular}{lll}
	\toprule
	Electric power systems  & Water distribution & Traffic control \\
	\midrule
	\makecell[l]{
		\textbullet\ \textit{Smart grids}: \citet{GungorSahinEtAl2011},\\ \citet{SinghKishorSamuel2016},\\ \citet{Antoniadou-PlytariaKouveliotis-LysikatosGeorgilakis2017}; \vspace{0.1cm}\\
		\textbullet\ \textit{Renewable energy integration}:\\
		\citet{TungadioSun2019},\\ \citet{MarinescuGomis-BellmuntEtAl2022},\\\citet{RanjanShankar2022},\\ \citet{DoerflerGros2023}.
	}
	& 
	\makecell[l]{
		\textbullet\ \textit{Irrigation}:\\
		\citet{Mareels2005irrigation}, \citet{CantoniWeyerLiEtAl2007},\\
		\citet{negenborn2009distributed}, \citet{heyden2022structured},\\ \revM{\citet{CobbenhagenSchoonenEtAl2022}}; \vspace{0.1cm}\\
		\textbullet\ \textit{Urban distribution}:\\
		\citet{MartinezEtAl2007}, \citet{NazifEtAl2010},\\ \citet{ocampo2009improving,ocampo2012hierarchical}.
	}
	& 
	\makecell[l]{
		\textbullet\ \textit{Urban signal control}:\\ \citet{AboudolasPapageorgiouKosmatopoulos2009},\\ \citet{AboudolasGeroliminis2013},\\
		\citet{PedrosoBatista2021}, \citet{TuetschHeEtAl2024};\vspace{0.1cm}\\		
		\textbullet\ \textit{Freeway network control}:\\ \citet{KotsialosPapageorgiouEtAl2002};\vspace{0.1cm}\\	
		\textbullet\ \textit{\revm{Connected} and automated vehicles}:\\\citet{KhayatianMehrabianEtAl2020},
	\citet{NielsMitrovicEtAl2020,NielsBogenbergerEtAl2024}.
	}\\
	\bottomrule
\end{tabular}
\end{table*}

Despite the \revm{potential} resilience, flexibility, and scalability of \rev{systems-of-systems}, and their extensive applicability across many diverse fields, most of the given examples are:
\begin{enumerate}[(a)] \itemsep-0em 
	\item \rev{Yet} to transition from conceptualization to deployment;
	\item \rev{Deployed} only in very controlled environments as a proof of concept; 
	\item \rev{Implemented} in practice on a very small scale with a small number of \revM{subsystems}.
\end{enumerate}
\revM{The reason is clear:}
\begin{mdframed}[style=callout,linecolor=blue]
	There are inhibiting technological challenges, especially regarding the feasibility of the implementation of state-of-the-art algorithms \revm{for the design and working phases of ultra} large-scale \rev{systems-of-systems}.
\end{mdframed}

Despite the large number and diverse nature of these emerging applications, three stand out due to their significance towards the sustainable development goals \citep{UN} called for by the United Nations (UN). These are:
\begin{itemize}\color{black}\itemsep0em
	\item \textcolor{black}{\revm{Ultra} large-scale agriculture;}
	\item \textcolor{black}{\rev{\revm{Ultra} large-scale electric power systems with spatially} distributed renewable sources;}
	\item \textcolor{black}{Navigation and orbit control of LEO mega-constellations.}
\end{itemize}
\revM{We will detail these further in the next subsections.}

\subsection{\revm{Ultra} Large-Scale Agriculture}

The world's population is projected to increase from 8 billion in 2022 to 9.7 billion in 2050 \citep{worldpop2022}. Thus, to meet the increasing demand, agricultural production in 2050 will need to yield almost 50\% more food, \revM{livestock} feed, and biofuel than it did in 2012 \citep{fao2017}. However, present-day resource intensive farming paradigms are unsustainable and cannot be scaled due to their contribution to climate-change, pressure on natural resources, and deforestation. Indeed, the whole food system is responsible for a third of the global greenhouse gas emissions, of which 71\% are directly connected with agriculture and land-use activities and the remainder with supply chain activities~\citep{crippa2021food}. As a result, it is now well established that the key to the sustainable agricultural growth is to develop more efficient  large-scale practices for the use of land and resources through technological progress \citep{OECD2011,fao2017}. Accordingly, it is one of the main challenges recently identified in the panel session ``Control for Societal-Scale Challenges: Road Map 2030" \citep{annaswamy2023control}, which is instrumental to strive towards the UN's Sustainable Development Goals~2 and~6  \citep{UN}.

In response to this challenge, agricultural R\&D is surging and the shift towards a smart agriculture paradigm seems inevitable \citep{liu2020industry,yang2021survey}. At the center of this evolution is digitization, featured by the integration of sensors, actuators, and \revm{ULS} coordination. Progress on this topic is mainly moving towards precision agriculture techniques, which rely on measuring and acting on temporal-spatial differences of agronomic measures to trade-off yield and economical and ecological costs~\citep{mcbratney2005future}. For example, crop growth is very sensitive to the rate and timing of fertilization. Thus, measuring the spacial-temporal evolution of agricultural indicators and carrying out precise fertilizer application is essential to maximizing their effectiveness \citep{tremblay2010development,cobbenhagen2021opportunities}.  The scalability of precision agriculture techniques procedures is enabled by: 
\begin{itemize} \itemsep0em
	\item The rise of robotics in agriculture \citep{bechar2016}, specifically \revm{ULS} swarms of ground \revm{and} aerial \rev{vehicles for crop monitoring}, water stress evaluation, chlorophyll density estimation, seeding, and fertilizer and pesticide spraying \citep{Radoglou-GrammatikisEtAl2020};
	\item The management of resources with spatially distributed sinks and \rev{sources, such as irrigation} \revm{networks}.
\end{itemize}

\subsection{\revm{Ultra} Large-Scale Power Grids}

The transition towards renewable energy resources is pressing and clear. With electricity demand expected to increase steadily in Europe at a compound annual growth rate of about 2\% until 2035, and increasing energy prices \citep{SamsethEtAl2021}, this shift is more urgent than ever. One of the big challenges of this transition is the integration of renewable energy resources in the \rev{electric} grid, since it requires complex large-scale distributed coordination, information processing, optimization, estimation, and monitoring \rev{\citep{DoerflerGros2023}}. \rev{However, as it is pointed out in \citet{StoustrupChakraborttyEtAl},} the current smart grid technology is not adequate for the increased use of \rev{spatially} distributed renewable sources, which are promptly becoming a major component of the power grid. These sources are uncertain, \rev{since} it is difficult to model the evolution of wind or solar power, which leads to increased power imbalances and frequency deviations in the network. Thus, without proper coordination schemes for \revm{ULSS}, the much-wanted high penetration of renewable energy sources may lead to a degradation of the power quality and distribution efficiency and, ultimately, instability. Therefore, the development of control algorithms for \revm{ULS} \rev{electric power systems} is crucial to facilitate the transition towards renewable sources and decarbonization \citep{SinghKishorSamuel2016,Antoniadou-PlytariaKouveliotis-LysikatosGeorgilakis2017}. Not only is it pointed out as one of the main societal-scale control challenges for 2030 \citep{annaswamy2023control}, but it is also called for in Goal 7 of the UN's Sustainable Development Goals \citep{UN}.


\subsection{Navigation and Control for LEO Mega-constellations }


The conceptualization of large LEO constellations began in the 1990s, with Globalstar, Iridium, Odyssey, and Teledesic, as an attempt to provide communication services globally. Nevertheless, none of these projects but Iridium \rev{succeeded. The} reason for their collapse was the reduced demand for these services and the high deployment costs, which rendered these projects inviable \citep{DaehnickKlinghofferMaritzEtAl2020}. Despite that, the technological advances and soaring demand for broadband connectivity that were witnessed in the following two decades have led to a reawakening of LEO mega-constellation projects \rev{namely~(i)~Telsat’s Telesat Lightspeed; (ii)~OneWeb; (iii)~SpaceX’s Starlink; and (iv)~Amazon’s Project Kuiper.} Large-scale constellations of satellites in LEO are, with the current technology, unarguably a solution to meet the increasing demand for reliable low-latency, high capacity, global broadband connectivity, which is in line with Goal 9 of the UN's sustainable development goals \citep{UN}. Nevertheless, there \rev{is still one substantial hurdle to the success of the new generation of LEO constellation, which is their economical viability \citep{DaehnickKlinghofferMaritzEtAl2020,ZhanEtAl2020}.}





The ongoing shift is undeniable, from the use of constellations of a small number of highly complex satellites to the employment of \revm{an ultra} large number of smaller and simpler satellites that cooperate in \revm{ULS} networks. In fact, the term mega-constellation has been coined to designate these \revm{ultra} large-scale constellations. For instance, the Starlink constellation is projected to feature 42000 satellites in LEO, of which over 5000 satellites have already been launched. Nevertheless, the aforementioned paradigm change has not yet been accompanied by a paradigm change from an \revm{operational} standpoint. As pointed out in \cite{ZhanEtAl2020}, the tracking, telemetry, and command (TT\&C) system projected for these constellations does not differ from the TT\&C system architecture employed for a single satellite. This system consists of a single centralized mission control center (MCC) with several ground terminals scattered across the globe to allow for continuous monitoring of the whole constellation. However, the implementation of such a centralized architecture \rev{for} a mega-constellation is very challenging and expensive, \rev{due to} its large dimension. The development of on-board control solutions is a pressing breakthrough in that direction. However, inter-satellite communication and computational and memory resources on-board each satellite are very limited, which imposes severe requirements on the design of a control solution that can be feasibly implemented.

\section{Objectives and Technological Challenges}\label{sec:challenges}

\revm{In this section, we resort to an abstract modeling setup to pinpoint and analyze the challenges that arise from an ULS.}

\subsection{Formulation and Objectives}\label{sec:fomulation}
In this section, we present \revp{a possible} \revm{overarching \revp{formulation}} \rev{to model the systems-of-systems under consideration in this paper. We resort to a graph representation whose nodes represent individual dynamical \revM{subsystems} and whose \revM{edges} represent interactions between them. This representation is \revM{useful for modeling a broad range of complex networks (see, e.g.,} \cite{BoccalettiLatoraEtAl2006} for a detailed analysis). \revM{In addition,} it is worth characterizing the interconnections between \revM{subsystems} on different layers, namely:
\begin{itemize}	\itemsep0em 
	\item Dynamics;
	\item Sensor output;
	\item Performance output;
	\item Communication.
\end{itemize}
The resulting model is a multilayer network of dynamical systems (see, \revM{e.g.,} \cite{BoccalettiBianconiEtAl2014,KivelaeArenasEtAl2014} for an analysis of \revM{such} models). This modeling framework was chosen for its generality and its versatility in formulating the requirements for a feasible implementation on \revm{an \revm{ULS}}. Alternative modeling frameworks for systems-of-systems include dynamic graphs~\citep{Siljak2008}, and, \revM{when} resorting to a state-space discretization, discrete-event systems~\citep{RudieLafortuneEtAl2003,TabuadaPappas2006,CassandrasLafortune2008} and decentralized discrete abstractions \citep{BoskosDimarogonas2019b,BoskosDimarogonas2019}, which also allow to frame logic specifications (see, for example, \cite{PolaDiBenedetto2019} for more details). A detailed analysis of \revM{(these and other)} alternative modeling approaches is outside the scope of this paper.}

\rev{Furthermore, a \emph{discrete-time} model of the dynamics of the individual \revM{subsystems} is considered, since it is closer in nature to the digital implementation of a controller. Nevertheless, an equivalent analysis can also be performed for a continuous-time model of the dynamics of the individual \revM{subsystems}, leading to an overall sampled-data system.}


\subsubsection{\rev{Formal} modeling setup}\label{sec:mod_setup}

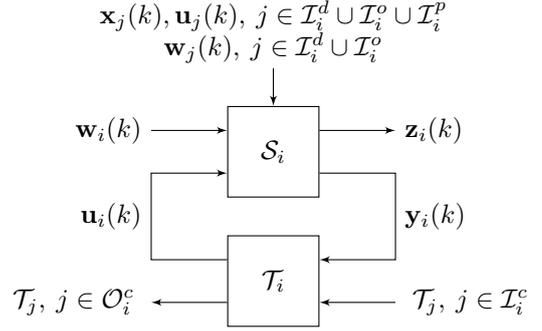
\begin{figure}[t]
   	\centering
	\begin{tikzpicture}[auto, node distance=1cm,>=latex']
			\node [block] (Si) {$\mathcal{S}_i$};
			\node [block, below=of Si, yshift=+.5cm] (Ti) {$\mathcal{T}_i$};
			\node [left=of Si.155] (wi) {$\revM{\mathbf{w}_i}(k)$}; 
			\node [left=of Si.205] (ui) {};
			\node [right=of Si.25] (zi) {$\revM{\mathbf{z}_i}(k)$};
			\node [above=of Si,yshift=-.5cm, label={[align=center, yshift=-0.3cm] above:$\revM{\mathbf{x}_j}(k), \revM{\mathbf{u}_j}(k),\: j\in \mathcal{I}_i^d \cup  \mathcal{I}_i^o \cup  \mathcal{I}_i^p$ \\$\revM{\mathbf{w}_j}(k),\: j\in \mathcal{I}_i^d \cup  \mathcal{I}_i^o$}] (xj) {}; 
			\node [left=of Ti.205] (Tj-in) {$\mathcal{T}_j,\: j\in \mathcal{O}_i^c\;$};
			\node [right=of Ti.335] (Tj-out) {$\;\mathcal{T}_j, \:j\in \mathcal{I}_i^c$};
			\draw[->] (wi) -- (Si.155);
			\draw[->] (Si.25) -- (zi);
			\draw[<-] (Tj-in) -- (Ti.205);
			\draw[<-] (Ti.335) -- (Tj-out) ;
			\draw[->] (xj) -- (Si) ;
			\draw[->] (Ti.155) -- ++ (-1,0) |- node [pos=0.25] {$\revM{\mathbf{u}_i}(k)$}  (Si.205) ;
			\draw[->] (Si.335) -- ++ (1,0) |- node [pos=0.25] {$\revM{\mathbf{y}_i}(k)$}  (Ti.25) ;

			
	\end{tikzpicture}    
	\caption{Block diagram \rev{of actuated \revM{subsystem} $\mathcal{S}_i$.}}
	\label{fig:sys_blk_draft}
\end{figure}

Consider a network of $N$ \revM{subsystems} $\mathcal{S}_i$, \revM{$i = 1,2,\ldots,N$}, each \rev{(possibly)} associated with one computational unit $\mathcal{T}_i$, which carries out the computation of the control input of $\mathcal{S}_i$ \rev{and handles communication with \revM{computational units of other \revm{subsystems}}}. Each \revM{subsystem}  $\mathcal{S}_i$, \revM{$i = 1,2,\ldots,N$}, is modeled by a generic discrete-time state-space system, which \revM{involves}, at time $k\in \mathbb{N}_0 = \{0,1,\ldots\}$, an input vector $\revM{\mathbf{u}_i}(k)\in \mathbb{R}^{m_i}$, an exogenous disturbance input $\revM{\mathbf{w}_i}(k)\in \mathbb{R}^{q_i}$, an internal state $\revM{\mathbf{x}_i}(k)\in \mathbb{R}^{n_i}$, a sensor output $\revM{\mathbf{y}}_i(k) \in \mathbb{R}^{o_i}$, and a performance output $\revM{\mathbf{z}_i}(k) \in \mathbb{R}^{p_i}$. \rev{Three particular flavors of \revM{subsystems} are of interest:
	\begin{enumerate}[(a)]\itemsep0.3em
	\item \emph{Actuated}: \revM{An actuated} \revM{subsystem \revm{has actuation capabilities} and is equipped with a computational unit that computes its control input} (possibly) resorting to communication with other \revM{subsystems} (e.g., a satellite in a constellation); 
	\item \emph{Cooperative Unactuated}: \revM{A cooperative unactuated} \revM{subsystem} is equipped with a computational unit to \revM{transmit information to} other \revM{subsystems} \revM{but does not have a control input} (e.g., an unactuated leader in a leader-following mission);
	\item \emph{Uncooperative Unactuated}: \revM{A uncooperative unactuated} \revM{subsystem} \revM{is} not equipped with a computational unit, and, thus \revM{does not transmit information to} other \revM{subsystems}, \revM{and does not have a control input} (e.g., space debris or an asteroid in a rendezvous maneuver). \revM{Note that it is still possible that the sensor or performance output of another subsystem has an interconnection with an uncooperative unactuated subsystem (e.g., a probe taking a radar measurement of the distance between itself and an asteroid).}
\end{enumerate}}%
\revp{A block diagram of an actuated subsystem $\mathcal{S}_i$ is depicted in Fig.~\ref{fig:sys_blk_draft}. Block diagrams for cooperative unactuated and uncooperative unactuated subsystems follow from removing the input and both the input and the computational unit, respectively, in Fig.~\ref{fig:sys_blk_draft}.}


\revM{We proceed to model the interconnections between subsystems on different layers.} First, the dynamic couplings can be represented by a directed graph $\mathcal{G}_d =(\mathcal{V},\mathcal{E}_d)$, which may be time-varying, composed of a set $\mathcal{V}:=\{\revM{1,2,\ldots,N}\}$ of vertices and a set $\mathcal{E}_d$ of directed edges.  In this framework, each system is represented by a vertex, i.e., \revM{subsystem} $\mathcal{S}_i$ is represented by vertex $i \in \mathcal{V}$. An edge $e$ incident on vertices $i$ and $j$, directed from $j$ towards $i$, is denoted by $e = (j,i)$. For a vertex $i$, its in-neighborhood $\mathcal{I}^d_i$ is the set of vertices from which there is an edge in $\mathcal{E}_d$ directed towards $i$. If the dynamics of $\mathcal{S}_i$ depend on the state or input of $\mathcal{S}_j$, then this coupling is represented by an edge directed from vertex $j$ towards vertex $i$ in $\mathcal{G}_d$, i.e., edge $e = (j,i) \in \mathcal{E}_d$. A self-loop in every vertex $i$ encodes the dependence of the dynamics of \revM{subsystem} $\mathcal{S}_i$ on its own state and input. It is important to stress that the direction of the edge matters. Note, for instance, that the fact that the dynamics of $\mathcal{S}_i$ depend on the state or input of $\mathcal{S}_j$ does not necessarily imply the converse. 

Second, each \revM{subsystem} has its own sensor and performance output signals, which may be coupled with the state or input of a set of other \revM{subsystems}. Similarly to the dynamic couplings, the sensor and performance output couplings can also be represented by directed graphs $\mathcal{G}_o = (\mathcal{V},\mathcal{E}_o)$ and $\mathcal{G}_p = (\mathcal{V},\mathcal{E}_p)$, respectively. To be clear, if $(j,i)\in \mathcal{E}_o$ ($(j,i)\in \mathcal{E}_p$) and, thus, $j\in \mathcal{I}^o_i$ ($j\in \mathcal{I}^p_i$), then the sensor (performance) output of $\mathcal{S}_i$ depends on the state or input of $\mathcal{S}_j$.

Third, each \revM{subsystem} may be able to directly establish a directed communication link with other \revM{subsystems} to receive information that is useful to compute its control action. Therefore, analogously, a directed communication graph $\mathcal{G}_c = (\mathcal{V},\mathcal{E}_c)$ may also be defined. In this case, since each computational unit always has prompt access to its own memory with negligible delay without resorting to any physical communication links, no self-loops are included in $\mathcal{G}_c$.


\revm{Each} \revM{subsystem} $\mathcal{S}_i$ can be generically modeled by the \revM{discrete-time} system
\begin{equation}\label{eq:localNonLinDynamics}
	\begin{cases}
		\revM{\mathbf{x}_i}(k+1) \!\!\!\!&= \revM{\mathbf{f}_i}\left(k;\revM{\mathbf{x}_j}(k),\revM{\mathbf{u}_j}(k),\revM{\mathbf{w}_j}(k); j\in {\mathcal{I}_i^d} \right)\\
		\revM{\mathbf{y}_i}(k)\!\!\!\! &=  \revM{\mathbf{g}_i}\left(k,\revM{\mathbf{x}_j}(k),\revM{\mathbf{u}_j}(k),\revM{\mathbf{w}_j}(k); j\in {\mathcal{I}_i^o} \right)\\
		\revM{\mathbf{z}_i}(k) \!\!\!\!&=  \revM{\mathbf{h}_i}\left(k,\revM{\mathbf{x}_j}(k),\revM{\mathbf{u}_j}(k); j\in {\mathcal{I}_i^p} \right),\\
	\end{cases}
\end{equation}
where functions $\revM{\mathbf{f}_i}\!:\!\mathbb{N}_0\!\times\!\prod_{j\in {\mathcal{I}_i^d}} \!\mathbb{R}^{n_j} \!\times \!\prod_{j\in{\mathcal{I}_i^d}}\! \mathbb{R}^{m_j} \times\prod_{j\in{\mathcal{I}_i^d}}\! \mathbb{R}^{q_j} \! \to \!\mathbb{R}^{n_i}$, $\revM{\mathbf{g}_i}:\mathbb{N}_0\times\prod_{j\in {\mathcal{I}_i^o}} \mathbb{R}^{n_j} \times\prod_{j\in{\mathcal{I}_i^o}} \mathbb{R}^{m_j} \times\prod_{j\in{\mathcal{I}_i^o}} \mathbb{R}^{q_j} \! \to \!\mathbb{R}^{o_i}$, and  ${\revM{\mathbf{h}_i}:\mathbb{N}_0 \times \prod_{j\in {\mathcal{I}_i^p}}\mathbb{R}^{n_j} \times \prod_{j\in {\mathcal{I}_i^p}}\mathbb{R}^{m_j} \!\to\! \mathbb{R}^{p_i}}$ are known multivariate functions that model the dynamics, sensing output, and performance output, respectively, of \revM{subsystem} $\mathcal{S}_i$. The control objective is to design a controller for each \revM{subsystem} $\mathcal{S}_i$, described by \eqref{eq:localNonLinDynamics}, that is implemented in $\mathcal{T}_i$ and maps the sensor output $\revM{\mathbf{y}_i}(k)$ to the control input $\revM{\mathbf{u}_i}(k)$ given information exchanged with other computational units $\mathcal{T}_j, j\in \mathcal{I}^c_i$, such that a given global metric
\begin{equation}\label{eq:global_cost}
	J := \sum\nolimits_{i = 1}^{N} J_i(\revM{\mathbf{z}_i}(\tau),\revM{\mathbf{u}_i}(\tau); \tau \in \mathbb{N}_0 )
\end{equation} 
is minimized, where $J_i$ is a local component that depends on $\revM{\mathbf{z}_i}(\tau)$ and $\revM{\mathbf{u}_i}(\tau)$ over a finite or infinite time window.




\begin{mdframed}[style=example,frametitle={\rev{\textbf{Example}: Satellite Constellation (running example)}}]
	
	To illustrate the \rev{concepts above}, consider a very simple illustrative running example, whose \rev{diagram} is depicted in Fig.~\ref{fig:eg_sat}, of the \revM{formation control} task of a constellation of satellites on a single orbital plane. The network-wise objective, \revM{to be expressed in \eqref{eq:global_cost},} is to maintain the \revM{nominal} separation between consecutive satellites. In this example, each satellite is represented by a system that controls its own thruster input. The dynamics of the satellites are decoupled, hence, in the dynamical graph $\mathcal{G}_d = (\mathcal{V},\mathcal{E}_d)$, $\mathcal{E}_d$ is simply composed of a self-loop in every vertex, where each vertex represents one satellite. We consider that position and velocity measurements are locally available to each satellite resorting to GNSS signals. Therefore, the sensor output graph is $\mathcal{G}_o = (\mathcal{V},\mathcal{E}_o)$, where $\mathcal{E}_o$ is composed of only a self-loop in every vertex. 
	 
	 {	\vspace{0.6cm}
	 	\centering
	 	\includegraphics[width = 0.75\linewidth]{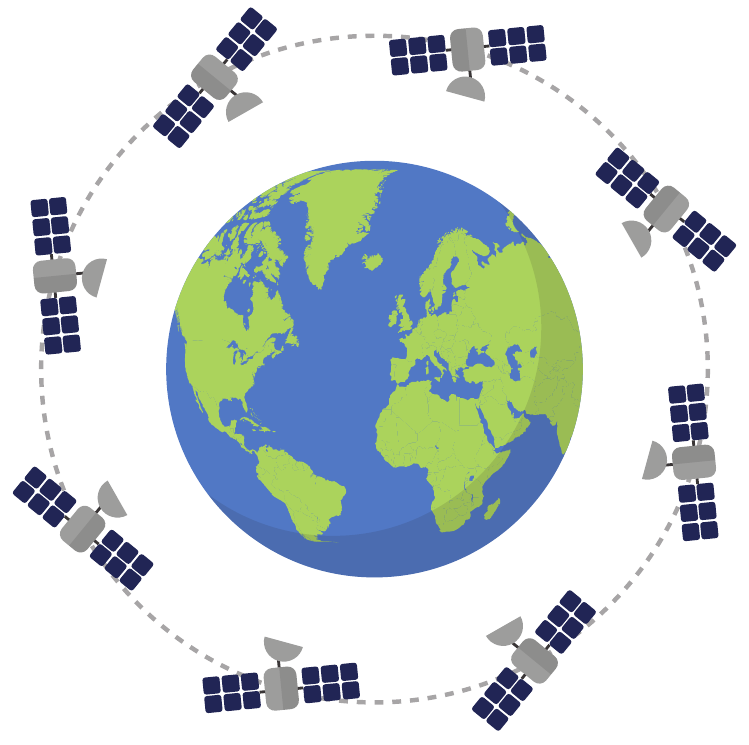}
	 	\captionof{figure}{Illustrative example of \rev{the} \revM{formation control} \rev{task} of a constellation of satellites on a single orbital plane (not to scale).}\label{fig:eg_sat}
	 	\vspace{0.6cm}
	 	\par}

	 We can define the performance output of each satellite as a two-dimensional vector of the angular spacing between itself and the preceding and succeeding satellites, whose reference is the nominal angular separation. Thus, the performance output coupling graph is $\mathcal{G}_p = (\mathcal{V},\mathcal{E}_p)$, where $\mathcal{E}_p$ is composed of edges directed towards every satellite from preceding and succeeding satellites and a self-loop in every satellite. We consider that \revM{the computational unit of} each satellite can exchange \rev{information} with the \revM{computational units of the} preceding and succeeding satellites. \rev{The nature} \rev{of the data exchanged depends on the cooperative control algorithm that is implemented.} Thus, the communication graph is $\mathcal{G}_c = (\mathcal{V},\mathcal{E}_c)$, where $\mathcal{E}_c$ is composed of edges directed towards every satellite from preceding and succeeding satellites. These four graphs define the topology of the \revM{system-of-systems} of this illustrative example, of which a section is schematically depicted in Fig.~\ref{fig:ill_network}.
	
	{	\vspace{0.6cm}
		\centering
		\includegraphics[width = 0.85\linewidth]{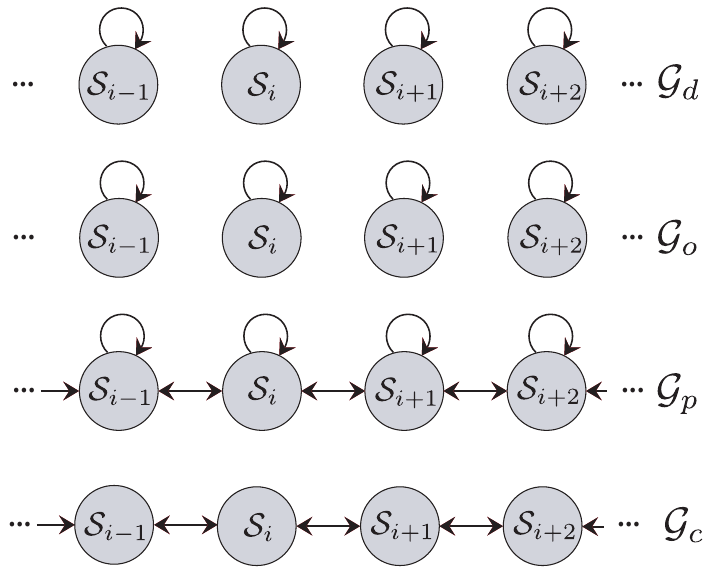}
		\captionof{figure}{Topology \rev{of the \revM{formation control} task of a constellation of satellites, depicted for a fraction of the satellite network in Fig.~\ref{fig:eg_sat}. The four layers of interactions: (i)~dynamics; (ii)~sensor output; (iii)~performance output; and (iv)~communication are depicted separately and according to the topology described.}}\label{fig:ill_network}
		\vspace{0.6cm}
		\par}
	
	
	To illustrate the formalization in \eqref{eq:localNonLinDynamics}, we assume a noise-free model for each \revM{subsystem} that can be described by linear time-varying \revM{dynamics}. We also consider that full-state feedback is available, i.e., $\revM{\mathbf{g}_i}(k,\revM{\mathbf{x}_i}(k),\revM{\mathbf{u}_i}(k)) = \revM{\mathbf{x}_i}(k)$ for all $i = \revM{1,2,\ldots,N}$. A \revM{subsystem} $\mathcal{S}_i$ is, then, described by 
	\begin{equation*}
		\begin{cases}
			\revM{\mathbf{x}_i}(k+1) \!\!\!\!&= \revM{\mathbf{A}_i}(k)\revM{\mathbf{x}_i}(k) + \revM{\mathbf{B}_i}(k)\revM{\mathbf{u}_i}(k) \\
			\revM{\mathbf{y}_i}(k)\!\!\!\! &=  \revM{\mathbf{x}_i}(k)\\
			\revM{\mathbf{z}_i}(k) \!\!\!\!&=  \sum_{j\in \mathcal{I}^p_i} \revM{\mathbf{H}_{ij}}(k)\revM{\mathbf{x}_j}(k),
		\end{cases}
	\end{equation*}
	where $\revM{\mathbf{A}_i}(k)$, $\revM{\mathbf{B}_i}(k)$, and $\revM{\mathbf{H}_{ij}}(k), j\in \mathcal{I}^p_i,$ at time $k$ are time-varying matrices of appropriate dimensions that model \revM{subsystem} $\mathcal{S}_i$. The control objective could be, for example, to design a controller such that a quadratic cost is minimized over an infinite window, i.e., 
	\begin{equation*}
		\sum_{i=1}^{N} \underbrace{\sum_{\tau = 0}^{\infty}\! \left(( \revM{\mathbf{z}_i}(\tau) \!-\! \revM{\mathbf{z}_i}^{\mathrm{ref}})\!^\top\!\revM{\mathbf{Q}_i}( \revM{\mathbf{z}_i}(\tau) \!-\! \revM{\mathbf{z}_i}^{\mathrm{ref}}) \!+\!  \revM{\mathbf{u}_i}\!^\top\!(\tau)\revM{\mathbf{R}_i}\revM{\mathbf{u}_i}(\tau)\right)}_{J_i(\revM{\mathbf{z}_i}(\tau),\revM{\mathbf{u}_i}(\tau); \tau \in \mathbb{N}_0)},
	\end{equation*}
	where $\revM{\mathbf{z}_i}^{\mathrm{ref}} \in \mathbb{R}^2$ encodes the reference mean argument of latitude separation between preceding and succeeding satellites and $\revM{\mathbf{Q}_i}$ and $\revM{\mathbf{R}_i}$ are positive semi-definite and positive definite matrices, respectively, of appropriate dimensions that weigh the tracking error and magnitude of the control action. The formulation of this example can be seamlessly extended, with the proposed general framework, to also account for partial sensing outputs, both measurement and process noise, and nonlinear dynamics, sensing outputs, and performance outputs as in \eqref{eq:localNonLinDynamics}.

\end{mdframed}

\begin{mdframed}[style=example,frametitle={\rev{\textbf{Example}: Estimation problem}}]
Interestingly, estimation problems can be cast within \revM{the framework of Section~\ref{sec:mod_setup} as well}. To make this clear, consider a network of $N$ \revM{subsystems} $\mathcal{S}_i$, $i = \revM{1,2,\ldots,N}$, with $\mathcal{G}_d = (\mathcal{V},\mathcal{E}_d)$, $\mathcal{G}_o = (\mathcal{V},\mathcal{E}_o)$, and $\mathcal{G}_c = (\mathcal{V},\mathcal{E}_c)$, which follow from the dynamics, sensing outputs, and communication graph in place, respectively, defined analogously to the generic model in Section~\ref{sec:mod_setup}. Denote the estimate of the state of \revM{subsystem} $\mathcal{S}_i$ at time $k$ by $\revM{\mathbf{\hat{x}}_i}(k)$. Taking the performance output of each \revM{subsystem} $\mathcal{S}_i$ for the estimation task as the state estimation error, i.e., $\revM{\mathbf{z}_i}(k) = \revM{\mathbf{x}_i}(k)-\revM{\mathbf{\hat{x}}_i}(k)$, the performance output graph $\mathcal{G}_p = (\mathcal{V},\mathcal{E}_p)$ is simply  composed of a self-loop in every vertex.  Then, the goal is to design an estimation solution to be deployed in $\mathcal{T}_i$ that maps the sensor output $\revM{\mathbf{y}_i}(k)$ to the state estimate $\revM{\mathbf{\hat{x}}_i}(k)$ given information exchanged with other computational units $\mathcal{T}_j, j\in \mathcal{I}^c_i,$ such that a given global metric $J := \sum_{i = 1}^{N} J_i(\revM{\mathbf{z}_i}(\tau); \tau \in \mathbb{N}_0 )$ is minimized. As an illustration, such a metric could be the average state estimation squared error over an infinite window, i.e.,   
\begin{equation*}
	J := \sum_{i=1}^{N} \lim_{T\to\infty} \frac{1}{T}\sum_{\tau = 1}^{T}\mathrm{tr}\left( \mathbb{E}_\mathbf{w}[\revM{\mathbf{z}_i}(\tau)\revM{\mathbf{z}_i}(\tau)^\top]\right),
\end{equation*}
where $\mathbb{E}_{\mathbf{w}}[\cdot]$ denotes the expectation operator, taken over the process and sensor noise distributions of \rev{$\revM{\mathbf{w}_j}, j =\revM{1,2\ldots,N}$}.
\end{mdframed}

\rev{Recall, from Section~\ref{sec:introduction}, that the control problem for systems-of-systems can be divided into two phases:~(i)~\revM{the \emph{design phase}}; and (ii)~\revM{the \emph{working phase}}. Given the formal modeling setup previously presented, these two phases can be defined more rigorously, \revp{for a standard \revd{information-constrained} control problem without any \revd{specific} ULS requirements \revd{(for now)},} as follows.

\begin{problem}[Design Phase]\label{prob:design}
	Given a priori available knowledge about:
	\begin{itemize}\itemsep -.1em%
		\item Graphs \revM{$\mathcal{G}_d, \mathcal{G}_o,\mathcal{G}_p, \mathcal{G}_c$} and functions \revM{$f_i, g_i, h_i$, $i\in \{1,2,\ldots,N\}$} that model the plant according to \eqref{eq:localNonLinDynamics};$\!\!\!$
		\item The disturbances;
		\item Mission specific design requirements (DR) (e.g., state and input constraints);
	\end{itemize}
	design, for each $i \in \{\revM{1,2,\ldots,N}\}$, a \revm{local} controller \revd{that maps}
	\begin{itemize}\itemsep-0.1em
		\item $\revM{\mathbf{y}_i}(k)$ and information from $\mathcal{T}_j,\: j\in \mathcal{I}_i^c$\vspace{-0.2cm}
	\end{itemize}
	\revd{to}
	\begin{itemize}\itemsep-0.1em\vspace{-0.2cm}
		\item $\revM{\mathbf{u}_i}(k)$ and information to $\mathcal{T}_j,\: j\in \mathcal{O}_i^c$
	\end{itemize}
	that minimizes the global cost \eqref{eq:global_cost} subject to the DR.
\end{problem}

\begin{problem}[Working Phase]\label{prob:working}
	\revd{For} each $i \in \{\revM{1,2,\ldots,N}\}$, at each discrete time \revM{$k\in \mathbb{N}_0$},  given a posteriori knowledge about \revd{$\mathbf{y}_i(k)$},  and information from $\mathcal{T}_j,\: j\in \mathcal{I}_i^c$, implement the local \revm{controller} designed in Problem~\ref{prob:design}, to \revd{provide} output $\revM{\mathbf{u}_i}(k)$ and information to $\mathcal{T}_j,\: j\in \mathcal{O}_i^c$.
\end{problem}

\revd{For Problem~\ref{prob:design} to be well-defined, the time evolution of the graphs $\mathcal{G}_d, \mathcal{G}_o,\mathcal{G}_p, \mathcal{G}_c$ and functions $f_i, g_i, h_i$, $i\in \{1,2,\ldots,N\}$ may need to be perfectly known (or predicted) into the future.} Intuitively, one can interpret these two phases in \revm{the} light of the block diagram in Fig.~\ref{fig:sys_blk_draft}. Indeed, notice that the \emph{design phase} amounts to optimally \revp{\emph{designing}} a control system \revm{composed of local controllers} for blocks $\mathcal{T}_i$, $i = \revM{1,2,\ldots,N}$, subject to the DR and the \emph{working phase} amounts to \revp{\emph{implementing}} the \revm{local controllers} in blocks $\mathcal{T}_i$, $i = \revM{1,2,\ldots,N}$, in real time.

\subsection{Challenges \revm{and Requirements}}\label{sec:challenges-subsec}

\revp{By} constraining the information structure resorting to the communication graph $\mathcal{G}_c$, the control solution \revd{implemented in the working phase} is \emph{distributed} (\revM{or} \emph{decentralized}, if $\mathcal{E}_c = \emptyset$). Indeed, the burden of computing the control \revd{actions} in the working phase is distributed among the \revM{subsystems}.}

\begin{mdframed}[style=callout,linecolor=blue]
		\rev{\revm{We} argue that the inevitable transition from large to \revm{ultra} large scale calls for \emph{additional requirements} \revd{beyond standard information constraints} to enable \revp{the} implementation of the design and working phases in real-life applications.}
\end{mdframed}

\rev{These additional requirements are necessary on four levels: 
	
\begin{itemize}\itemsep0.3em
	\item \emph{Topological}: \revm{The} number of interconnections (on any layer) of a \revM{subsystem} \revm{cannot grow} unbounded \revM{as $N\to \infty$}. \revm{Otherwise,} the amount of information to keep track of in \revp{one or both of the phases} grows unbounded \revM{as $N\to \infty$};
	\item \emph{Design}: \revm{The design phase must be distributed across subsystems.} State-of-the-art design procedures for \revM{computing the} distributed \revM{controller in the design phase (almost always)} \revm{rely} on a centralized computation. As $N$ increases (from large to \revm{ultra} large scale), the design procedure eventually becomes infeasible to be carried out in a single computational entity, especially if it has to be performed \revM{online};
	\item \emph{Computational and memory}: The local computational and memory resources in each computational unit necessary for performing \revm{both phases} cannot grow unbounded as  \revM{$N\to \infty$};
	\item \emph{Communication}: \revm{The} number of communications links that each \revM{subsystem} establishes with other \revM{subsystems} cannot grow \revM{unbounded as $N \to \infty$}. \revm{Otherwise,} the memory and computational burden of the \revM{\emph{working phase}} grows unbounded \revM{as $N \to \infty$}. \revm{Moreover, the data transmissions cannot be considered instantaneous.}
\end{itemize}}

\rev{The requirements above focus strongly on establishing that the quantity of local resources required in each \revM{subsystem} \revm{to perform both phases} \revM{is bounded as $N \to \infty$}. The nature of these requirements follows seamlessly from the concept of \emph{scalability} of systems-of-systems introduced in Section~\ref{sec:introduction}. Indeed, as more \revM{subsystems} are deployed (e.g., to increase throughput of the mission) the resources that are required locally in each of the (already deployed) \revM{subsystems} \revM{do not increase above a certain bound}. \revM{If the systems are initially deployed with resources capable of handling such bound, then the integration of more subsystems is seamless.}}

\begin{mdframed}[style=prospects,frametitle={\revm{\textbf{Remark}: Learning-based control}}]
	\revm{In learning-based control solutions, the \revd{division into} design and working phases \revd{is not as obvious}.  The overall learning-based procedure, which is implemented in each computational unit, \revd{can be perceived as the working phase and it} must still follow the aforementioned topological, computational, memory, and communication requirements.}
\end{mdframed}


In what follows, in Sections~\ref{sec:topological}--\ref{sec:communication}, \rev{the aforementioned} requirements are detailed thoroughly, which are all then summarized in Table~\ref{tab:req}, and some further guidelines are also provided in Section~\ref{sec:guidelines}. To ease the presentation of these \revm{requirements}, they are illustrated for the running satellite constellation example, \rev{which is described in Section~\ref{sec:mod_setup} and depicted in Figs.}~\ref{fig:eg_sat} and~\ref{fig:ill_network}. Henceforth, big $\mathcal{O}$ notation is employed to express bounds on asymptotic growth. Specifically, we say that a function $f:\mathbb{N}\to\mathbb{R}$ grows with $\mathcal{O}(g(n))$, where $g:\mathbb{N}\to\mathbb{R}_{\geq 0}$, if \revm{there are an $M\in \mathbb{R}_{\geq 0}$ and an $n_0\in  \mathbb{N}$ such that for all $ n \geq n_0$ it holds that $|f(n)| \leq Mg(n)$.}

\subsubsection{Topological Requirements}\label{sec:topological}

As mentioned above, the overall topology of the control problem is represented by four (possibly time-varying) directed graphs regarding the dynamical couplings, sensor and performance output couplings, and communication links that are established. In this setting, two vertices are coupled if there is an interaction of some nature between the corresponding \revM{subsystems}. It is important to point out that the topology of the control problem does not merely stem from the nature of the system, but also from \revm{planning choices (e.g., formulation of the mission requirements and choice of sensors and actuators fitted on the subsystems)}. Specifically, the dynamical, sensor output, and performance output topologies reflect the nature of the system and \revm{planning} choices that stem from it. Yet, the communication topology is purely a \revm{planning} choice that, \revm{given the choice of sensors and actuators,} trades off financial and deployment feasibility on the one hand and performance and resilience on the other. In this section, we focus on \revm{topological} requirements on dynamical, sensor output, and performance output topologies. Requirements on the communication topology are addressed in Section~\ref{sec:communication}.
	

\rev{Interactions between \revM{subsystems} in \revp{typical} systems-of-systems of \revm{an \revm{ULS}} are local. Thus it is natural to assume that:}
\begin{itemize}\itemsep0.3em
	\item The \textcolor{black}{dynamical graph has a local topology, in the sense that the number of \revM{subsystems} with which a \revM{subsystem} has dynamic couplings does not scale with the number of systems in the whole network, i.e., $|\mathcal{I}_i^d|$ grows with $\mathcal{O}(1)$ \revm{w.r.t.}\ the number of \revM{subsystems} in the network \revm{$N$};}
	\item The \textcolor{black}{sensing and objective design choices are such that the sensor output and performance output graphs are local, in the sense that the number of \revM{subsystems} with which a \revM{subsystem} has sensor or performance output couplings does not scale with the number of \revM{subsystems} in the whole network, i.e., $|\mathcal{I}_i^o|$ and $|\mathcal{I}_i^p|$ grow with $\mathcal{O}(1)$ \revm{w.r.t.}\ \rev{$N$}.}
\end{itemize}

Note that these requirements are \rev{intrinsic to meaningful systems-of-systems of \revm{an \revm{ULS}}.  To} illustrate this, consider \revm{the} two examples \rev{previously \revm{described} in Section~\ref{sec:applications}}. 

\begin{mdframed}[style=example,frametitle={\rev{\textbf{Example}: Satellite constellation (running example)}}]
	\revm{Consider again} the satellite constellation example described in Section~\ref{sec:mod_setup}. Notice that the dynamical, sensor output, and performance output graphs of this example, which are depicted in Fig.~\ref{fig:ill_network}, abide by the aforementioned requirements. 
\end{mdframed}

\begin{mdframed}[style=example,frametitle={\rev{\textbf{Example}:  Irrigation network}}]
	Consider \rev{an example of an} irrigation network, whereby each \revM{subsystem} is an open-air canal whose state is the water-level and whose inputs are the flows from each canal that is directly upstream. The dynamics of \rev{canal} $i$ are coupled with the state and input of a \rev{canal} $j$ that is directly upstream. As a result, there is an edge directed from $j$ to $i$ in $\mathcal{G}_d$. The sensor output is the local water-level in each canal, and the network-wise control objective is to track a set-point water-level in each canal, thus $\mathcal{G}_o$ and $\mathcal{G}_p$ are just composed of self-loops in every vertex \citep{negenborn2009distributed}. Again, notice that $\mathcal{G}_d$, $\mathcal{G}_o$, and $\mathcal{G}_p$ abide by the aforementioned topological requirements.
\end{mdframed}



\subsubsection{\rev{Design} Requirements}\label{sec:synthesis}

\rev{The} overwhelming majority of the procedures for the \rev{design} phase of \rev{decentralized and distributed} control laws in the literature are performed on a single computational \rev{entity}, i.e., the \rev{\emph{design phase procedure is still central}}. \revp{Well-known central convex optimization approaches (e.g.,~\citet{BoydBarratt1991,SchererWeiland2000,BoydVandenberghe2004}) \revd{shift the boundary between large and ultra large scale, in the sense that systems of a larger scale can be handled in a central computational entity (and hence are not considered ULSS).}} \revp{Nevertheless, as $N$ increases, the design procedure eventually becomes infeasible to be carried out in a single computational entity.} \revd{This calls for distributed design procedures that subdivide the design problem into multiple smaller problems each with access to partial system information that is locally available.} \revd{Hence, in a system-of-systems, whereby each subsystem is equipped with a computational unit:}
	
\begin{mdframed}[style=callout]
		\revd{An ULS calls for distributed design procedures that distribute the design phase among the computational units of the subsystems.}
\end{mdframed}

\revp{Historically}, concerns about the scalability of the \rev{design} phase of decentralized control solutions have been voiced long ago \citep{siljak1996}. This aspect has subsequently been \revp{partially undertook} in the literature by developing \revp{procedures that rely \revd{on} parallel computations across multiple computational cores}, e.g., \citet{siljak2005}. At first sight, it may appear that state-of-the-art parallel procedures for the \rev{design} phase \revp{generally} translate directly to distributing it \revp{among computational units \revd{on} an ULS in a scalable manner}. However, \rev{that is far from being the case.} Indeed, \revp{computational units \revd{in ULSS have limited} computational and memory resources}, and the communication links between \revp{them} are few and subject to limitations such as bandwidth constraints and delays, \revd{which are key concerns not typically taken into account} for parallelizing computations. In Section~\ref{sec:sota}, the state-of-the-art in \rev{decentralized and distributed} control is \revm{outlined} with more detail, making this fact clearer.  

Moreover, if each \revM{subsystem} in the network has time-invariant dynamical, sensor output, and performance output models, time-invariant \revm{topologies}, and time-invariant network-wise control \revm{objectives}, one can usually perform the \rev{design} phase centrally prior to deployment, which is often designated as \emph{offline}. In this case, the \rev{design} phase only has to be carried out once, which \revp{also \revd{shifts} the boundary between large and ultra large scale.} However, in more \revp{typical} scenarios, \revp{the} dynamical, sensor output, or performance output models, the network topology, or control objective are time-varying and cannot be predicted accurately beforehand. \revp{For example:}

\begin{mdframed}[style=callout]
\revp{Deploying additional (or recalling already deployed) subsystems, which is the premise behind the potential scalability of systems-of-systems, requires that the control solutions are able to accommodate these changes. This \revd{is related with the notions of \emph{open multi-agent systems} \citep{HendrickxMartin2017,AbdelrahimHendrickxEtAl2017}} and of  \emph{plug-and-play capability} \citep{Stoustrup2009}. In this case, the design phase has to be carried out regularly \revd{online} and distributing the design phase among the subsystems is even more compelling.}
\end{mdframed}






%
%
%



\revp{These points have} also recently been \revm{briefly pointed out specifically for industrial cyber-physical systems in} \cite{DingHanWangEtAl2019}, calling suitable \rev{\emph{distributed {design} procedures}} that account for resource limitations. In what follows, \revm{such} resource constraints are analyzed further and formulated formally.

\begin{mdframed}[style=example,frametitle={\rev{\textbf{Example}: Satellite constellation (running example)}}]
	Figs.~\ref{fig:ill_synthesis_centralized} and~\ref{fig:ill_synthesis_distributed} \rev{depict schemes of a centralized and a distributed design} architecture, respectively, for the \rev{satellite constellation running} example. Instead of focusing all the computational load associated with the \rev{design} phase on a single computer, which must transmit and receive data from all satellites, a distributed architecture allows for a distributed computation on local small computational units associated with each satellite. As the dimension of the constellation reaches \revm{an \revm{ULS}}, the communication and computational burden of the centralized architecture in Fig.~\ref{fig:ill_synthesis_centralized} becomes infeasible to implement in practice and the transition to a distributed \rev{design} procedure, \revm{as schematically depicted in Fig.~\ref{fig:ill_synthesis_distributed},} is inevitable. 
	
	{	\vspace{0.6cm}
		\centering
		\includegraphics[width = 0.9\linewidth]{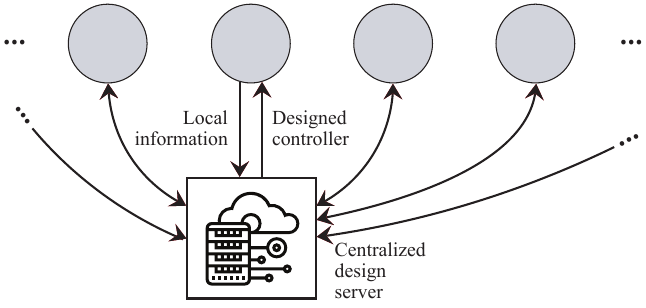}
		\captionof{figure}{\rev{Scheme of a centralized design} for the illustrative network in Fig.~\ref{fig:eg_sat}. \rev{Local information is shared from each \revM{subsystem} to the central computational entity, which centrally computes a design procedure in real time. The local controllers are then \revm{deployed} to each of the \revM{subsystems}, which locally implement the working phase.}}\label{fig:ill_synthesis_centralized}
		\vspace{0.4cm}
	\par}
	
		{	\vspace{0.2cm}
		\centering
		\includegraphics[width = 0.95\linewidth]{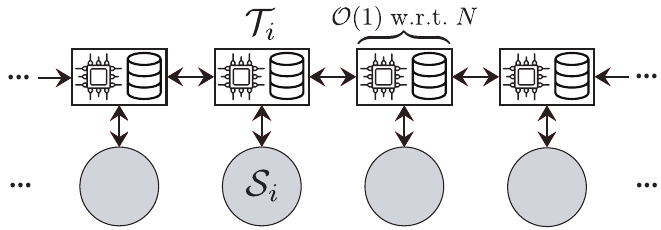}
		\captionof{figure}{\rev{Scheme of a distributed design} for the illustrative network in Fig.~\ref{fig:eg_sat}. \rev{The burden of the design phase is shared among the \revM{subsystems} relying on locally available information and communication with neighboring \revM{subsystems}. The working phase is also \revm{carried out} locally in each \revM{subsystem}. \revm{The asymptotic} computational and memory resources in each computational unit must not scale w.r.t.\ $N$.}}\label{fig:ill_synthesis_distributed}
	\vspace{0.2cm}
		\par}
\end{mdframed}

\begin{table*}[h!]
	\centering
	\caption{Requirements for a feasible implementation \revd{on} \revm{an \revm{ULS}}, which are thoroughly described in Section~\ref{sec:challenges}.}
	\label{tab:req}
	\vspace{-0.1cm}
	\small
	\begin{tabular}{p{2cm}p{0.5cm}p{12cm}}
		\toprule
		Category & \# & Requirements \\
		\toprule
		\rev{Design} & 1 & Distributed \rev{design} phase\\ 
		\midrule
		Computational & 2 & The local computational complexity grows with $\mathcal{O}(1)$ \revm{w.r.t.}\ $N$\\
		\midrule
		Memory & 3 & The local memory complexity grows with $\mathcal{O}(1)$ \revm{w.r.t.}\ $N$\\
		\midrule
		Communication & 4.1 & The local communication complexity grows with $\mathcal{O}(1)$ \revm{w.r.t.}\ $N$ \\
		& 4.2 & Hard real-time transmissions are not allowed (only applies to synchronous protocols)  \\
		\bottomrule
	\end{tabular}
\end{table*}


\subsubsection{Computational and Memory Requirements}

Computational and memory resources available in the computational unit of each \revM{subsystem} are often limited due to factors such as cost, weight, and energy. Therefore, the computational load of the real-time implementation of the \revm{design and working phases} must be distributed across all \revM{subsystems} in such a way that the replication of computations among \revM{subsystems} is reduced to a minimum. \rev{Indeed}, the computational and memory capabilities of each \revM{subsystem} should not be tailored to a particular network size, which would inhibit the seamless integration of additional \revM{subsystems} and severely hinder scalability. For these reasons:
\begin{itemize}\color{black} \itemsep0.3em
	\item  The \textcolor{black}{computational complexity of the floating-point operations carried out in the computational unit associated with each \revM{subsystem} must not scale with the number of \revM{subsystems} in the whole network;}
	\item  The \textcolor{black}{amount of data to be stored in each \revM{subsystem} must not scale with the number of \revM{subsystems} in the whole network.}
\end{itemize}
These limitations can be expressed as bounds on the asymptotic growth of \revm{computational and memory usage} as the dimension of the network increases: \rev{The} computational and memory complexity of \rev{each} \revM{subsystem} grows with $\mathcal{O}(1)$ w.r.t.\ $N$. 

\vspace{0.3cm}

\begin{mdframed}[style=example,frametitle={\rev{\textbf{Example}: Satellite constellation (running example)}}]
	\revm{The computational and memory requirements are} very compelling in the satellite constellation \rev{running} example in Fig.~\ref{fig:eg_sat}. Indeed, the computational and memory resources available on-board satellites are naturally very limited. On top of that, consider that the constellation of this example is upgraded and the number of satellites is doubled. To seamlessly support the upgrade, the computational and memory burden of the originally deployed satellites must not grow. These asymptotic requirements are also depicted in Fig.~\ref{fig:ill_synthesis_distributed} for the illustrative running satellite example.
\end{mdframed}

\vspace{0.3cm}

\subsubsection{Communication Requirements}\label{sec:communication}

In this section, we define the communication requirements. To do so, we \revm{assume} that all \revM{subsystems} have \emph{synchronized clocks} and information transmission is allowed periodically with a given period $T$ along the edges of the communication graph. In practice, this synchronization can be achieved with a GNSS receiver if high clock synchronization accuracy is required, or simpler technology otherwise.

\rev{First, to} obtain scalable communication infrastructure, \rev{a requirement} is that the number of available communication links between \revM{subsystems} is limited. To seek a scalable solution, the number of \revm{communication} links established with each \revM{subsystem} must not scale with the number of \revM{subsystems} in the whole network. Define, similarly to \citet{KiaRoundsEtAl2014}, the communication complexity of a certain \revM{subsystem} as the number of communication links that, at a given instant, are established with other \revM{subsystems}. Thus, the local communication complexity at the protocol level must grow with $\mathcal{O}(1)$ w.r.t.\ $N$.

Second, the amount of data transmitted per \revM{subsystem} must not scale with the number of \revM{subsystems} in the whole network. This constraint comes as a consequence of the bounds on asymptotic growth of the memory and communication complexities.

Third, in synchronous control solutions, data transmission cannot be considered \rev{to be} instantaneous. \revm{Consider a computation in $\mathcal{T}_i$ at time instant $k$ that requires a variable computed in  $\mathcal{T}_j$ at the same time instant. In this case, the completion of the computation in $\mathcal{T}_i$ depends on the processing delay in $\mathcal{T}_j$  and transmission delay of data from $\mathcal{T}_j$, which requires very accurate (and thus complex) synchronization algorithms and fast processing and transmission hardware. These type of transmissions are} denoted as \rev{synchronous} hard real-time transmissions. \revm{Conversely, consider a computation in $\mathcal{T}_i$ at time instant $k$ that requires a variable computed in $\mathcal{T}_j$ at the time instant $k-1$. When there is a processing delay of $\Delta_c$ time units in $\mathcal{T}_j$ and a transmission delay of $\Delta_t$ time units, the information can be made available in $\mathcal{T}_i$ at time $t = (k-1)T+ \Delta_c +\Delta_t$.  These type of transmissions are denoted as synchronous soft real-time transmissions.}

\begin{mdframed}[style=callout,linecolor=blue]
	 \revm{Soft real-time} transmissions are robust to transmission and computation delays that are small compared to the transmission period, and do not require clock synchronization as accurate as hard real-time transmissions. 
\end{mdframed}
\begin{mdframed}[style=example,frametitle={\rev{\textbf{Example}: Satellite constellation (running example)}}]
	
	In Fig.~\ref{fig:oill_soft_hard_com}, hard real-time transmissions are depicted for the satellite \rev{constellation running} example, for a time instant $k$ and considering computation and communication delays of $\Delta$ time units. \revm{Notice that Fig.~\ref{fig:oill_soft_hard_com} depicts hard real-time transmissions since, for instance, the operation $\phi_j$ in $\mathcal{T}_i$ at time instant $k$ requires $\phi_j(y_j(k))$, which is only computed in $\mathcal{T}_j$ at the same time instant.} \rev{From Fig.~\ref{fig:oill_soft_hard_com}, it} is clear that, in solutions that rely on synchronous protocols, hard real-time transmissions cannot be feasibly implemented \revd{on} \revm{an \revm{ULS}} due to the accumulation of delays.
	
	{\vspace{0.3cm}
		\centering
		\includegraphics[width = \linewidth]{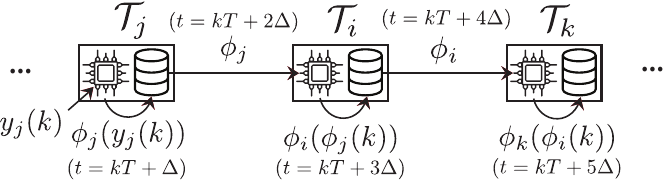}
		\captionof{figure}{\rev{Scheme depicting delays} resulting from \rev{synchronous} hard real-time communications with period $T$ for the satellite example in Fig.~\ref{fig:eg_sat}. \revm{The computational} unit $\mathcal{T}_j$ has access to a measurement $y_j(k)$ at time $kT$. For the sake of the illustration, we consider that it takes $\Delta$ time units to evaluate some function of $y_j(k)$, e.g., $\phi_j(y_j(k))$, and also $\Delta$ time units for that result to be transmitted to computational unit $\mathcal{T}_i$. Computational unit $\mathcal{T}_i$ also performs an operation $\phi_i$ on $\phi_j(y_j(k))$, \revm{which takes} $\Delta$ to be computed and $\Delta$ to be received in $\mathcal{T}_k$. \rev{As a result, that} value is only ready to be used in $\mathcal{T}_k$ at time $kT+4\Delta$.	\label{fig:oill_soft_hard_com} }
	\vspace{0.4cm}
	\par}

\revm{Moreover, the use of soft real time transmissions does not allow sharing} data through a path of established communication links via other \revM{subsystems} in one transmission instance. \revm{Still,} the transmission of information between two \revM{subsystems} through a path of $M$ other \revM{subsystems} is possible throughout $M+1$ transmission instances, reaching the receiving \revM{subsystem} with a \revm{nominal} delay of \revm{$M$ time steps}.
	
\end{mdframed}

\subsubsection{Further Guidelines}\label{sec:guidelines}

The requirements at the \rev{design}, topological, computational, memory, and communication level presented up to this point are crucial for \revm{a feasible} implementation on \revm{an \revm{ULS}}. Nevertheless, two important guidelines are also worth pointing out \revm{as they}, although not imperative or relevant in all applications, may contribute significantly to the cost effectiveness of control solutions over \revm{ULS} networks:
\begin{itemize}\itemsep0.3em\color{black}
	\item  \emph{Asynchronous communication}: Network-wide \textcolor{black}{synchronous communication without access to an external common signal requires very complex synchronization protocols that are very challenging to implement. Therefore, if the \revM{subsystems} are not equipped with a GNSS receiver or similar synchronization \revm{technologies}, the \revm{design and working phases} must rely on asynchronous communication and sensor measurements. In that case, the control problem is significantly more involved, and the nominal \revm{discrete-time} dynamics of each \revM{subsystem} have to be \revm{obtained} with a time-varying discretization interval.}
	\item \emph{Resource-aware strategies}: Sensing \textcolor{black}{couplings are very costly in harsh environments, since the measurement itself requires energy and often additional communication.  Similarly, updating the control action often requires additional energy and the exchange of information between \revM{subsystems}. This is particularly relevant in applications whereby energy consumption is the bottleneck of the lifespan of the mission, or the available communication channels have low bandwidth and high loss rate. This is the case, for example, \rev{in} underwater applications \citep{zereik2018challenges}. Consequently, measurements and control action updates can be carried out only whenever necessary, instead of periodically, making a compromise between performance and energy and communication \rev{load}. These \rev{trade-offs} should be exploited for efficiency and may be employed in real time making use of, for instance, self-  and event-triggered control and communication techniques \citep{heemels2012introduction,nowzari2019event}.}
\end{itemize}

\section{State-of-the-art Overview}\label{sec:sota}

With the emergence of \revm{ULSS}, there are several technical challenges that hamper their real-life application, which \revm{were} outlined in Section~\ref{sec:challenges}. In this section, the most popular state-of-the-art decentralized and \rev{distributed} estimation and control approaches are detailed in \revm{the} light of their technical difficulties in overcoming these challenges, with emphasis on the possibility of a distributed \rev{design phase}.

Decentralized \rev{and distributed} estimation and control \revm{have} been one of the most researched topics in control theory over the past decades, resulting in a \revm{great} number of important contributions. The surveys and books \citet{lunze1992feedback,siljak2005,bakule2008,Siljak2012,Bakule2014,OhParkAhn2015,DingHanWangEtAl2019,EspinaLlanosBurgos-MelladoEtAl2020,KordestaniSafaviSaif2021} offer overviews of this extensive topic. \rev{\revm{The} \emph{design} phase of a distributed control problem, on which a substantial research effort has been put forth since early work in the sixties \citep{mcfadden1969controllability}, is, in general, intractable.} Even for linear time-invariant (LTI) systems, the problem of designing controllers and estimators, which consists \revm{of} solving an optimization problem subject to a (generic) sparsity constraint that arises from the \rev{distributed} nature of the configuration, is extremely difficult, as discussed in \citet{BlondelTsitsiklis2000}, and remains an open problem. In fact, the optimal solution for a linear system with Gaussian noise may be nonlinear, as pointed out in \cite{Witsenhausen1968}. Furthermore, it has been shown that the solution of a \rev{distributed} design control problem is the result of a convex optimization problem, and thus tractable, if and only if quadratic invariance of the controller set is ensured \citep{Rotkowitz2005,LessardLall2010,LessardLall2015}, which preserves the \rev{distributed} control structures under feedback. This condition is, unfortunately, \revm{rarely} verified. Ergo, more complex problems that involve time-varying dynamics or topologies or even nonlinearities are bound to be as (or even more) intractable. Analogously to classical estimation and control, \rev{distributed} estimation and control algorithms can be perceived as dual problems, which make use of similar techniques and \rev{design} procedures. Therefore, in what follows, the state-of-the-art techniques presented are suitable for the application to both problems unless otherwise mentioned.

\subsection{Consensus-based Approach}\label{sec:consensus}

\rev{First, consensus arguments have historically been one of the first approaches to explain and enforce cooperative behavior in multi-agent systems with decoupled dynamics \citep{BorkarVaraiya1982,TsitsiklisBertsekasEtAl1986}. In general, consensus algorithms use local communication between \revM{subsystems} to achieve \revm{a} network-wide agreement (consensus) on a common variable that is used for cooperation. Consensus arguments explain cooperation in nature between large numbers of animals \citep{SumpterKrauseEtAl2008} and can be employed to enforce cooperative behavior such as:
\begin{itemize}
	\item \emph{Flocking}, whereby dynamically decoupled \revM{subsystems} cooperatively achieve cohesion, separation, and alignment \citep{Reynolds1987} (e.g., \citet{JadbabaieLinEtAl2003,olfati2006flocking,CuckerSmale2007,SepulchrePaleyEtAl2008});
	\item \emph{Rendezvous}, whereby dynamically decoupled \revM{subsystems} cooperatively converge towards a single unspecified location (e.g., \citet{LinMorseEtAl2003,LinMorseEtAl2004});
	\item \emph{Formation control}, whereby dynamically decoupled systems cooperatively acquire specified positions relative to each other (e.g., \citet{FaxMurray2004,PorfiriRobersonEtAl2007,Ren2007});
	\item \emph{Sensor networks}, whereby a network of sensors reaches a consensus on the state of a common process that is observed (e.g., \citet{Olfati-SaberShamma2005,Olfati-Saber2007,CarliChiusoEtAl2008}).
\end{itemize}
Furthermore, consensus algorithms are endowed with very attractive qualities, such as great flexibility and low communication requirements. An overview of consensus algorithms and their properties can be seen in \revm{\citet{RenBeardAtkins2005,RenBeard2007,GarinSchenato2010,AmirkhaniBarshooi2022,NingHanEtAl2023}}. In general, consensus arguments provide a scalable methodological solution for control problems of \revm{ULSS} of dynamically decoupled \revM{subsystems}  \cite[Chap.~8]{RenBeard2007}. 

\vspace{0.6cm}

\begin{mdframed}[style=example,frametitle={\rev{\textbf{Example}: \revm{Ultra} large-scale consensus}}]
\rev{Consider a system of $N$ \revM{subsystems} with single integrator dynamics, i.e., \vspace{-0.2cm}
\begin{equation*}
	x_i(k+1) = x_i(k) + u_i(k), \quad i = \revM{1,2,\ldots,N},\vspace{-0.2cm}
\end{equation*}
a time-invariant undirected and connected communication graph $\mathcal{G}_c$, and a diffusive coupling input \begin{equation}\label{eq:consensus_u}
u_i(k) = -g_i\sum_{j\in \mathcal{I}_i^c}(x_i(k)-x_j(k)),  \quad i = \revM{1,2,\ldots,N}, \vspace{-0.1cm}
\end{equation}
where $g_i$ is the gain of each \revM{subsystem} $i$. If the only objective is to rendezvous, i.e., $|x_i-x_j|\to 0$ for all $i,j$ as $k\to \infty$, this solution is suitable for \revm{an ULS}:
\begin{enumerate}[(a)]
	\item \emph{Design phase}: If, in each \revM{subsystem} $i$, a positive gain ${g_i >0}$ is locally chosen, rendezvous is guaranteed \citep{Olfati-SaberMurray2004}. Therefore, it can be implemented satisfying the \revm{ULS} requirements.
	\item \emph{Working phase}: The control law \eqref{eq:consensus_u} can be implemented satisfying the \revm{ULS} requirements.
\end{enumerate}}\vspace{-0.3cm}
\end{mdframed}

Nevertheless, if one considers the problem of finding the \emph{optimal} consensus law, i.e., endowing the \emph{design phase} of a consensus-based control solution with a performance metric:
\begin{itemize}
	\item The problem is nonconvex in general \citep{JiaoTrentelmanEtAl2020};
	\item In particular cases where a closed-form solution is known, finding local control laws requires global information about the system \citep{CaoRen2010}.
\end{itemize}%
\revp{Therefore:}

\begin{mdframed}[style=callout,linecolor=blue]
	\rev{\revm{The} design phase of optimal consensus-based solutions is challenging to distribute among the \revM{subsystems} \revm{on an ULS}.}
\end{mdframed} \vspace{-0.2cm} Recent} efforts in this field to obtain a distributed \rev{design} procedure for LQ consensus \citep{Jiao2020,BarkaiMirkinEtAl2022,DijCha_CDC23a} \rev{are a promising starting point to enable optimal consensus-based solutions \revm{on an ULS}.}

\vspace{-0.2cm}

\subsection{$\mathcal{H}_2/\mathcal{H}_\infty$ BMI-based Approach}

\rev{Second}, one popular formulation for robust and optimal control and estimation of LTI systems with structural constraints inherited from the \rev{distributed} framework is to design an $\mathcal{H}_2$ or $\mathcal{H}_\infty$ optimal policy, which amounts to solving a bilinear matrix inequality (BMI), as presented, for instance, in \citet{ZuoNayfeh2003,ZhaiEtAl2006,BompartNollApkarian2007,ShahParrilo2013}. Nevertheless, solving BMIs is well-known to be NP-hard \citep{TokerOzbay1995}. For the centralized case, it is possible to use a variable change to obtain an equivalent convex problem that requires solving linear matrix inequalities (LMIs), for which there exist computationally efficient methods \citep{boyd1994linear,SchererWeiland2000}. However, the introduction of sparsity constraints on the gain (that portray the \rev{distributed} nature of the problem) precludes the use of a variable substitution since, in general, it does not preserve the sparsity constraint. Most of the results on the general case focus on using different methods to derive suboptimal solutions for the BMI, for instance \citet{ScorlettiDuc1997,ZhaiEtAl2006,BompartNollApkarian2007}. Moreover, suboptimal solutions for the BMI can be found \citep{Siljak2012} when considering specific decomposition structures such as overlapping decomposition and border block diagonal decomposition, which cover a very wide class of relevant control problems. This approach is extended for linear time-varying (LTV) systems in \citet{FarhoodDiDullerud2015} and for time-varying network topologies in \citet{ViegasEtAl2015}. In \citet{LinFardadJovanovic2013} and \citet{lopez2014sparse}, instead of a sparsity constraint, a sparsity inducing term is added to the BMI, yielding sparsity patterns that are a good compromise between performance and the extent of decentralization and communication cost. 

\begin{mdframed}[style=example, frametitle={\rev{\textbf{Example}: Robust structured control}}]
	\rev{Consider a system of $N$ \revM{subsystems} with LTI dynamics, a communication graph $\mathcal{G}_c$, and a linear control law
	\begin{equation}\label{eq:local_control}
	\revM{\mathbf{u}_i}(k) = -\sum_{j\in \mathcal{I}_i^c}\revM{\mathbf{K}_{ij}}\revM{\mathbf{x}_j}(k),  \quad i = \revM{1,2,\ldots,N},
	\end{equation}
where $\revM{\mathbf{K}_{ij}}$ is a gain matrix of appropriate dimensions. Concatenating the states and inputs of all \revM{subsystems} we obtain global state and input vectors, denoted by $\mathbf{x}(k) \in \mathbb{R}^n$ and $\mathbf{u}(k)\in\mathbb{R}^m$, respectively. The distributed control law \eqref{eq:local_control} can be written as $\mathbf{u}(k) = -\mathbf{K}\mathbf{x}(k)$, where $\mathbf{K}$ is a block matrix subject to a structural constraint $\mathbf{K}\in S$, where
\begin{equation*}
	S = \{\mathbf{K} \;|\;  \revm{\forall i,j} \; j\notin \mathcal{I}_i^c \implies \revM{\mathbf{K}_{ij}} = \mathbf{0}\}.
\end{equation*}
For example, the optimal design w.r.t.\ the $\mathcal{H}_2$  norm can be expressed as
\begin{equation}\label{eq:He_synth}
	\begin{aligned}
		& \min_{\mathbf{K},\mathbf{P},\mathbf{Z},\gamma} & & \gamma \\
		& \quad \textrm{s.t.} & & \mathbf{X}(\mathbf{K},\mathbf{P},\mathbf{Z}) \succeq \mathbf{0} \\
		& & & \mathrm{trace}(\mathbf{Z})\leq \gamma \\
		& & & \mathbf{K} \in S
	\end{aligned}
\end{equation}
where $\mathbf{X}(\mathbf{K},\mathbf{P},\mathbf{Z})$ is a block matrix that is bilinear w.r.t.\ its arguments \rmr{such that $\gamma$ is an upper bound of the $\mathcal{H}_2$ norm of the closed-loop system} \citep{SchererWeiland2000}. Since solving \eqref{eq:He_synth}} \rev{for $\mathbf{K}$ requires solving \revm{a BMI}, the design phase requires global information and cannot be distributed among the \revM{subsystems}. Therefore, the design phase does not follow the requirements for an implementation on \revm{an ULS}.}
\end{mdframed}
\begin{mdframed}[style=callout,linecolor=blue]
	Although algorithms for the solution of LMIs and BMIs are well-known, their parallelization across multiple cores is challenging \citep{Ivanov2008}, and distributing their computation among the \revM{subsystems}, while abiding by the communication, computational, and memory \rev{requirements}, is not possible.
\end{mdframed}


\begin{table*}[ht]
	\centering
	\caption{ Characterization of the state-of-the-art techniques in relation to the challenging aspects of \revm{an ULS} implementation.}
	\label{tab:sota}
	\small
	\begin{tabular}{p{2cm} p{3.2cm} p{3.5cm} p{3.2cm} p{3.3cm}}
		
		\toprule
		Technique  & Scope  & \rev{Design} & Comp. and Mem. & Communication \\
		\toprule
		\rev{Consensus} & \rev{Time-varying topologies; Dynamically decoupled} & \rev{Nonconvex and/or requires global information}  & \rev{Local implementation of \rev{control law}} & \rev{Local; Synchronous / Asynchronous} \\
		\midrule
		$\mathcal{H}_2$/$\mathcal{H}_\infty$ BMI-based  &  Time-varying topologies; Linear dynamics  & Solving global BMIs or LMIs iteratively; Offline \rev{design} & Local implementation of \rev{control law} & Synchronous\\
		\midrule
		Convex $\quad \quad$ relaxation  &  Time-varying topologies; Linear dynamics & Global closed-form synthesis; Offline \rev{design}  & Local implementation of \rev{control law} & Synchronous\\
		\midrule
		
		Cooperative $\;$localization  & Time-varying topologies; Nonlinear dynamics& Distributed \rev{design} under approximation & Comp. and/or mem. requirements scale with $N$ & Asynchronous \\
		\midrule
		Clustering  & Weak couplings between clusters  & Distributed among clusters; Centralized cluster-wise & Scales with the dimension of the clusters & All-to-all cluster-wise; Synchronous\\
		\midrule
		Distributed $\quad$ MPC  & Time-varying topologies; Nonlinear dynamics & Distributed & Does not scale with $N$ & Commonly requires hard-real time transmissions \\
		\midrule
		Convex formulation  & Strict invariance conditions & Exact global convex synthesis & Local implementation of \rev{control law} & Synchronous\\
		
		\bottomrule
	\end{tabular}
\end{table*}

\subsection{Convex Relaxation}

\rev{Third}, due to the aforementioned intractability of the general problem, another approach is to attempt to derive a suboptimal linear estimator or feedback controller by approximating the original nonconvex optimization problem by a convex one, which allows to make use of well-known optimization techniques. This approach is called convex relaxation \citep{Jain_2017} and it has been successfully leveraged in the past to solve complex problems in control theory \citep{HedlundRantzer2002,PrajnaParriloEtAl2004,Low2014a,Low2014b}. One relaxation approach, for the particular case of output feedback of a LTI system subject to a fixed sparsity constraint, is employed in \citet{wang2014} and \citet{wang2018convex}. Another structured \rev{design} approach, presented in \citet{PedrosoEtAl2022} and \citet{PedrosoBatista2023}, is suitable for LTV systems and time-varying typologies. Such sparsity constraints impose certain entries of the global gain matrix to be null, following a structure that reflects the \rev{distributed} nature of the network. By considering the network as a whole, the couplings between \revM{subsystems} are not neglected, but the \rev{design phase cannot} be distributed \rev{among} the \revM{subsystems}. The \rev{latter} relaxation approach can also be formulated as a single optimization problem for LTI systems \citep{ViegasEtAl2018,ViegasEtAl2021}, but it cannot even be computed offline in parallel, since it relies on operations on global matrices.

\begin{mdframed}[style=example, frametitle={\rmr{\textbf{Example}: Distributed receeding horizon LQ control}}]
	\rmr{In this example, we briefly illustrate the convex relaxation approach in \cite{PedrosoBatista2023}. Consider a system of $N$ subsystems with LTV dynamics, a communication graph $\mathcal{G}_c$, and a linear control law
		\begin{equation}\label{eq:local_control_tv}
			\mathbf{u}_i(k) = -\sum_{j\in \mathcal{I}_i^c}{\mathbf{K}_{ij}(k)\mathbf{x}_j}(k),  \quad i = {1,2,\ldots,N},
		\end{equation}
		where ${\mathbf{K}_{ij}}(k)$ is a gain matrix of appropriate dimensions. Concatenating the states and inputs of all {subsystems} we obtain global state and input vectors, denoted by $\mathbf{x}(k) \in \mathbb{R}^n$ and $\mathbf{u}(k)\in\mathbb{R}^m$, respectively. The distributed control law \eqref{eq:local_control_tv} can be written as $\mathbf{u}(k) = -\mathbf{K}(k)\mathbf{x}(k)$, where $\mathbf{K}(k)$ is a block matrix subject to a structural constraint $\mathbf{K}(k)\in S$,} \rmr{where
		\begin{equation*}
			S = \{\mathbf{K} \;|\;  {\forall i,j} \; j\notin \mathcal{I}_i^c \implies {\mathbf{K}_{ij}} = \mathbf{0}\}.
		\end{equation*}
		Consider a receding horizon problem that at each time $k$ consists of an optimal control problem on a window into the future, which is given by 
		\begin{equation}\label{eq:non_cvx}
			\begin{aligned}
				& \min_{\substack{\mathbf{K}(\tau), \\\tau = k,\ldots,k+T-1}} & & J(k) \\
				& \quad \textrm{s.t.} & &\mathbf{K}(\tau) \in S, \;\; \tau = k,\ldots,k+T-1,
			\end{aligned}
		\end{equation}
		where $T\in \mathbb{N}$ is the window length and $J(k)$ is a quadratic cost that weights the states and inputs along the window. It is well-known in the optimal control literature \citep{AndersonMoore1990,LewisVrabieEtAl2012} that such quadratic cost with a linear  control law can be written as $J(k) = \mathbf{x}^\top(k)\mathbf{P}(k)\mathbf{x}(k)$. Matrix $\mathbf{P}(k)$ is positive semidefinite and is given by a closed-form backwards recursion that depends on the quadratic cost weighting parameters, the dynamics of the LTV system, and $\mathbf{K}(\tau), \tau = k,\ldots,k+T-1$. The optimization problem~\eqref{eq:non_cvx} is nonconvex and depends on $\mathbf{x}(k)$, which is challenging to solve efficiently. Instead of solving it, one can consider instead the relaxed problem
		\begin{equation}\label{eq:cvx}
			\begin{aligned}
				& \min_{\substack{\mathbf{K}(\tau), \\\tau = k,\ldots,k+T-1}} & & \mathrm{tr}(\mathbf{P}(k)) \\
				& \quad \textrm{s.t.} & &\mathbf{K}(\tau) \in S, \;\; \tau = k,\ldots,k+T-1.
			\end{aligned}
		\end{equation}
		Crucially, \eqref{eq:cvx} can be cast as a convex problem and admits a closed-form solution for the sequence of global gains. For more details and an analysis of the relation between \eqref{eq:non_cvx} and \eqref{eq:cvx} see \cite{PedrosoBatista2023}.}
\end{mdframed}
\vspace{-0.3cm}

\begin{mdframed}[style=callout,linecolor=blue]
	\rev{Although convex relaxation significantly improves the scalability of central design procedures, it is still challenging to distribute the design phase among the \revM{subsystems}. }
\end{mdframed}

\subsection{Cooperative Localization}

Fourth, a very interesting problem that falls into the \rev{distributed} estimation problem for large-scale \rev{systems-of-systems} is the cooperative localization problem \rev{\citep{FoxBurgardEtAl2000,RoumeliotisBekey2002}.} \rev{It refers to the task of locally estimating the state of each \revM{subsystem} resorting to (i)~relative measurements w.r.t.\ a group other \revM{subsystems} and/or beacons with a known state, and (ii)~local communication.} In \citet{DaiMuWu2016}, the application of a junction-tree-based protocol for the distributed implementation of the centralized extended Kalman filter (EKF) is proposed. In \citet{LeungBarfootLiu2009}, the cooperative localization architecture makes use of extensive measurement storage that scales with the size of the network. Notwithstanding the lack of loss of performance in relation to the centralized approach, the communication and memory burden, respectively, of these approaches is massive, which renders its application to a large-scale network infeasible. Another well-known approach is to approximate the covariance propagation to meet the requirements for a feasible implementation. However, this scheme is prone to become over-confident by underestimating the correlation between measurements, which is designated as double-counting. This effect is exemplified in \citet{PanzieriPascucciSetola2006}. Covariance intersection (CI) and split covariance intersection (SCI) methods have been proposed to mitigate double-counting, for instance in \citet{Carrillo-ArceNerurkarGordilloEtAl2013,WanasingheEtAl2014}, employing resources that \rev{do not scale with the number of \revM{subsystems} in the network under the assumption of a local sensor output topology.} Recently, in \citet{LuftEtAl2016,LuftEtAl2018}, a promising recursive decentralized method, that computes an approximate correlation distributively between each pair of \revM{subsystems} for a general network and supports asynchronous communication and measurements, is proposed. This solution achieves good performance, and \revm{the} communication burden that does not scale with the dimension of the network. Although it does not rely on measurement bookkeeping, the memory requirements scale linearly with the number of systems in the network \revm{and cannot provably avoid underestimating the correlation between measurements.}

\begin{mdframed}[style=callout,linecolor=blue]
 	\rev{There has been promising evolution towards cooperative localization strategies based on a covariance intersection approach that are suitable for \revm{an ULS}. Since these are very conservative, recent methods focus on achieving better performance. However, this is often obtained at the expense of requiring local resource usage that scales with the number of \revM{subsystems} in the network.}
\end{mdframed}

\subsection{Clustering-based Approach}\label{sec:clustering}

\revm{Another} approach found in the literature is to decouple the network of \revM{subsystems} into clusters of \revM{subsystems} and consider the interactions between distinct clusters as disturbances. Applying standard techniques to each of the clusters, \rev{resorting to intra-cluster communication,} allows to obtain a control law that can be implemented in a \rev{distributed} configuration and \rev{whose design can be carried out distributively among clusters}, \rev{without resorting to inter-cluster communication \citep{Ocampo-MartinezBovoEtAl2011}}. \revm{However,} unless the inter-cluster couplings \rev{are weak, which is often an assumption on the cluster partition \citep{ZhengWeiEtAl2018},} this approach \rev{may lead} to significant performance losses. \rev{To mitigate that effect, a hierarchical approach is often taken resorting to a \revm{multi-layer controller} that \revm{also} manages the cluster partition and/or mediates the exchange of information between clusters \citep{FeleMaestreEtAl2017}.} The \rev{survey} \citet{ChanfreutMaestreCamacho2021} extensively \rev{reviews} algorithms \rev{for the partition of systems into clusters. Briefly, clustering methods can be divided into three categories \citep{ChanfreutMaestreCamacho2021}:
\begin{itemize}\itemsep0em
	\item \emph{Coalitional}: Clusters are dynamically shaped by enabling/disabling communication between \revM{subsystems} to trade-off performance and coordination cost. Clusters are computed offline and switching between coalitions occurs online (e.g., \citet{MaestreMunozdelaPenaEtAl2014,ChanfreutMaestreEtAl2021});
	\item \emph{Community detection}: Clusters are statically formed making use of graph-theoretical properties (e.g., \citet{JogwarDaoutidis2017,Jogwar2019});
	\item \emph{Time-varying}: Clusters are centrally optimized online to trade-off performance and coordination cost in a time-varying environment (e.g., \citet{Barreiro-GomezOcampo-MartinezEtAl2019,AnandutaPippiaEtAl2019}). 
\end{itemize}}

\begin{mdframed}[style=example,frametitle={\rev{\textbf{Example}: Clustering}}]
	\rev{Consider an illustrative system-of-systems depicted in Fig.~\ref{fig:cluster}, where solid edges represent an interaction at the dynamic, sensor output, or performance output levels. The system is partitioned into three clusters, as depicted.}
	
	{	\vspace{0.4cm}
		\centering
		\includegraphics[width = .75\linewidth]{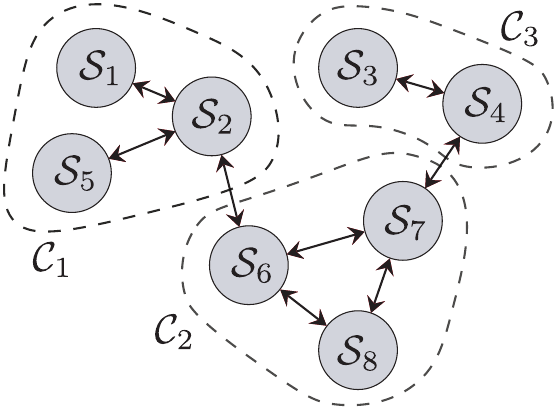}
		\captionof{figure}{\rev{Scheme of an illustrative clustering configuration.}}\label{fig:cluster}
		\vspace{0.4cm}
	\par}

	\rev{Given this clustering configuration, the simplest distributed control solution for each cluster $\mathcal{C}_i$, consists \revm{of}: 
	\begin{enumerate}[(a)]
		\item \emph{Design phase}: In each cluster $\mathcal{C}_i$, locally design a centralized control law for the concatenation of the \revM{subsystems} in $\mathcal{C}_i$, e.g., a linear quadratic regulator or a model predictive control scheme. The design may also rely on information exchanged with other clusters (e.g., $\mathcal{C}_1$ may receive the state of $\mathcal{S}_6$) to mitigate the performance loss due to neglecting inter-clustering couplings (e.g., the coupling between $\mathcal{S}_1$ and $\mathcal{S}_6$).
		\item \emph{Working phase}: In each cluster $\mathcal{C}_i$, implement the control law that was designed, resorting to intra-cluster communication between the \revM{subsystems}. 
	\end{enumerate}
	Notice that both the design and working phases are carried out locally and classical centralized control techniques can be employed. If the number of subsystems in each cluster does not grow with the dimension of the network, this approach is scalable to an ULS. \rmr{Moreover, there is a trade-off between communication and computational requirements (that scale with the cluster size) and performance. Such trade-off is analyzed for meaningful applications, for example, in \cite{Chanfreut2021,chanfreut2023aladin}.}}
\end{mdframed}
\vspace{0.4cm}
\begin{mdframed}[style=callout,linecolor=blue]
	\rev{\revm{Given} a \emph{precomputed clustering configuration}, the design and working phases of control algorithms can seamlessly be distributed among the clusters. That is achieved at the expense of performance, since couplings are neglected in the design phase, and all-to-all intra-cluster communication. If the clustering is performed dynamically to improve performance, then the online clustering procedure itself must follow the \revm{ULS} requirements. However, state-of-the-art dynamic clustering procedures are based on central procedures that require global system information \citep{ChanfreutMaestreCamacho2021}.}
\end{mdframed}

A clustering-based approach \revm{is} applied, for instance, \rev{to} the control of large-scale urban road networks \citep{ChowShaLi2019,Chanfreut2021}, large-scale smart grids \citep{BahramipanahCherkaouiEtAl2016,RahnamaBendtsenEtAl2017,HanTucciEtAl2019,LaBellaKlausEtAl2022}, and solar parabolic trough plants \citep{Chanfreut2023}.

\subsection{Distributed Model Predictive Control}\label{sec:DMPC}

Sixth, an additional approach worth noting is to employ distributed optimization tools to synthesize \rev{distributed} control solutions \citep{Nedic2018,testa2023tutorial}. Regarding this topic, the bulk of the attention is on the parallelization of model predictive control (MPC) \citep{Mayne2014,Rawlings2017}, denominated in the literature as distributed model predictive control (DMPC), whose concept and architectures are extensively detailed, for instance, in \cite{scattolini2009architectures,christofides2013distributed,muller2017economic}. DMPC approaches can be classified according to several criteria. A scheme is said to be:
\begin{enumerate}[(a)]\color{black}
	\item \emph{Cooperative} \textcolor{black}{if the local objective of each \revM{subsystem} portrays a network-wise objective, which is the setting under consideration in this paper;}
	\item  \emph{Iterative} \textcolor{black}{if, at each sampling instance, several iterations of communication and local computation must be performed.}
\end{enumerate}
It is immediate that \emph{iterative} DMPC schemes require hard real-time transmissions, and thus do not abide by the aforementioned \revm{ULS} requirements. Indeed, iterative DMPC schemes mainly stem from decomposition approaches based on distributed optimization, which require several iterations until \rev{an agreement} among the subproblems is achieved. Examples are \cite{Conte2016,chanfreut2023aladin,Jong2023}. Research in this direction focuses on speeding up convergence \citep{necoara2011parallel,giselsson2013accelerated} and on innovative parallelization strategies \citep{doan2011iterative}. 

\begin{mdframed}[style=callout,linecolor=blue]
	\revm{On an ULS}, the DMPC schemes of interest must be cooperative and non-iterative. Indeed, research into DMPC schemes that are cooperative and non-iterative is a subset of the family of problems that fall \rev{within} the scope of this paper. However, little research has been conducted on those schemes, and none of the works reviewed by \cite{scattolini2009architectures} are simultaneously cooperative and non-iterative.
\end{mdframed}

 Two of the few non-iterative cooperative schemes are \cite{RichardsHow2004,RichardsHow2004b}. Although of high interest, they rely on a sequential communication hierarchy to enforce network-wise constraints, which also requires hard real-time transmissions despite not being iterative. Another very interesting approach is proposed in \cite{KeviczkyBorrelliEtAl2006} for dynamically decoupled nonlinear \revM{subsystems}, which abides by all the \revm{ULS} requirements in Table~\ref{tab:req}. Therein, sufficient stability conditions are derived as a function of the mismatch between local solutions to the subproblems. But given a certain topology and dynamics of the network, a bound on this mismatch is not provided. Moreover, the global equilibrium has to be known. In this setting, if the mismatch between local solutions is bounded, they are said to be consistent (also sometimes called compatible) \citep{muller2017economic}. It is also important to mention the approaches in \cite{Dunbar2006} and \cite{Dunbar2007} which tackle the DMPC problems of dynamically decoupled nonlinear \revM{subsystems} with a network-wise objective and dynamically coupled nonlinear \revM{subsystems} with a decoupled objective, respectively. The topology of the network is time-invariant and the global equilibrium is time-invariant and known. Remarkably, the \revm{ULS} requirements in Table~\ref{tab:req} are also satisfied and, \rev{under} the assumption that there exists a stable decentralized time-invariant linear feedback terminal controller at the equilibrium, even consistency between local subproblems is achieved and stability can be guaranteed. In \cite{Zhao2015}, a similar approach as in \cite{Dunbar2006} is employed for a quasi-infinite horizon. More recently,  in \citet{PedrosoBatista2022} and \citet{PedrosoBatista2023b}, for the particular case of linear-time varying dynamically decoupled \revM{subsystems}, a DMPC scheme and an analogous EKF \rev{design} procedure, respectively, are proposed, relying on a convex relaxation procedure. Although they abide by the requirements in Table~\ref{tab:req} and are suitable for time-varying topologies, no consistency between subproblems or stability guarantees are provided.



\subsection{Convex Formulation for Special-structure Systems}

Seventh, for systems which verify the quadratic invariance condition detailed above, it is possible to achieve a convex, and thus tractable, formulation \citep{Scherer2013,Rotkowitz2005,LessardLall2015,LamperskiLessard2015}. Related approaches include optimal robust control over partially ordered sets \citep{ShahParrilo2013}, linear quadratic regulator design for the two-player problem \citep{SwigartLall2010}, and optimal control of discrete-time partially nested systems \citep{LamperskiDoyle2012}. The case of spatially invariant systems is treated in \citet{BamiehPaganiniDahleh2002,BamiehVoulgaris2005}, while \citet{Mahmoud2010,MahmoudAlmutairi2009} address systems with bounded nonlinear interconnections. Finally, powerful results can be found for even more specific classes of systems, such as positive systems \citep{TanakaLangbort2011,vladu2022decentralized} and symmetric Hurwitz state matrices \citep{LidstromRantzer2015}. 

\begin{mdframed}[style=callout, linecolor=blue]
	\revm{Approaches for special-structure systems} are very interesting from a theoretical standpoint, providing valuable \rev{insights} into the intricacies of the \rev{decentralized and distributed} control \rev{problems}. Nevertheless, the limiting assumptions on the system, imposed to achieve tractability, are rarely encountered in real-life applications. \revm{In addition,} although these particular schemes allow for a more efficient \rev{design phase}, eventually, as the scale increases, they also become infeasible to synthesize centrally.
\end{mdframed}

In Table~\ref{tab:sota}, a characterization of these state-of-the-art techniques in \revm{the} \rev{light of} the requirements \rev{for} \revm{an ULS} implementation is provided.


\section{Prospects}\label{sec:prospects}

\begin{table*}[ht]
	\centering
	\caption{Characterization of \rev{promising} state-of-the-art techniques in relation to the challenging aspects of \revm{an ULS} implementation.}
	\label{tab:prospects}
	\vspace{0.3cm}
	\small
	\begin{tabular}{c p{3.5cm} p{4cm} c c c c c  p{5cm}}
		\toprule
		&\multicolumn{1}{c}{\multirow{2}{*}{Works}} & \multicolumn{1}{c}{\multirow{2}{*}{Scope}} & \multicolumn{5}{c}{Requirements -- $\;$ Table~\ref{tab:req}}  & \multicolumn{1}{c}{\multirow{2}{*}{Guarantees}} \\ \cmidrule(lr){4-8}
		&&                        & 1 & 2 & 3 & 4.1 & 4.2  &                               \\
		\toprule
		
		\parbox[t]{2mm}{\multirow{4}{*}{\rotatebox[origin=c]{90}{\rev{Consensus$\;\;$}}}}
		& \rev{\citet{Olfati-SaberMurray2004}} & \rev{Single-integrator dynamics; Time varying topology} & \rev{\checkmark} & \rev{\checkmark}& \rev{\checkmark}& \rev{\checkmark}& \rev{\checkmark}& \rev{Consensus is achieved; Design does not optimize any performance metric}\\
		\cmidrule(lr){2-9}
		& \rev{\citet{Jiao2020,DijCha_CDC23a}} & \rev{Single-integrator dynamics; Performance-driven design} &\rev{\checkmark} &\rev{\checkmark} & \rev{\checkmark} & \rev{\checkmark} & & \rev{Consensus is achieved} \\
		\midrule
		
		\parbox[t]{2mm}{\multirow{6}{*}{\rotatebox[origin=c]{90}{\rev{$\;\;$Coop. Loc.}}}}
		& \rev{\citet{Carrillo-ArceNerurkarGordilloEtAl2013,WanasingheEtAl2014}}  &  \rev{Decoupled linear dynamics; Time-varying topology} &  \rev{\checkmark}  & \rev{\checkmark}    &  \rev{\checkmark}    &   \rev{\checkmark}  & \rev{\checkmark}   &  \rev{Provably consistent; Follows requirements under local topology assumption; Asynchronous communication}  \\
		\cmidrule(lr){2-9}
		& \citet{LuftEtAl2018} &  Decoupled linear dynamics; Time-invariant topology &  \checkmark  &  \checkmark  &     &   \checkmark    & \checkmark    &  No consistency guarantees; \rev{Asynchronous communication} \\
				
		\midrule
		
		\parbox[t]{2mm}{\multirow{1}{*}{\rotatebox[origin=c]{90}{\rev{Clusters}$\,$}}}
		& \rev{\citet{MaestreMunozdelaPenaEtAl2014,ChanfreutMaestreEtAl2021} } & \rev{Linear time-invariant dynamics; Hierarchical control: high level for dynamic clustering}  &    & \rev{\checkmark}     &  \rev{\checkmark}     &   \rev{\checkmark}   & \rev{\checkmark}    &    \rev{Stability guarantees}\\
		
		\midrule
		
		\parbox[t]{2mm}{\multirow{12}{*}{\rotatebox[origin=c]{90}{\rev{DMPC}}}}
		& \citet{RichardsHow2004} &  Decoupled linear dynamics; Leader-follower time-invariant topology &  \checkmark  &  \checkmark  &   \checkmark  &   \checkmark    &     &   Constraint satisfaction guarantees \\
		\cmidrule(lr){2-9}
		& \citet{KeviczkyBorrelliEtAl2006} &  Decoupled nonlinear dynamics; Time-varying topology; Global equilibrium has to be known  &  \checkmark  &  \checkmark  &   \checkmark  &   \checkmark    &   \checkmark  &     No consistency or stability guarantees \\
		\cmidrule(lr){2-9}
		& \citet{Dunbar2006,Dunbar2007} &  Nonlinear dynamics; Time-invariant topology; Global equilibrium has to be known  &  \checkmark & \checkmark  &  \checkmark  &   \checkmark  &  \checkmark &     Consistency and stability guarantees, given the existence of a time-invariant decentralized stabilizing controller \\
		\cmidrule(lr){2-9}
		& \citet{PedrosoBatista2023b,PedrosoBatista2022} &  Decoupled linear dynamics; Time-varying topology &  \checkmark  &  \checkmark  &   \checkmark  &   \checkmark    &   \checkmark  &    No consistency or stability guarantees  \\
		
		\bottomrule
	\end{tabular}
\end{table*}

\rmr{In Section~\ref{sec:sota}, the analysis of several state-of-the-art approaches in the light of the technical challenges that arise on an ULS allowed to draw insightful conclusions. Overall, we conclude that current techniques do not adequately address these challenges.} In this section, we provide insight into how state-of-the-art techniques can be leveraged to \revm{start addressing} this problem.

\subsection{\rev{Promising Techniques}}

\rev{A few families of techniques offer control solutions that, under simple settings, (partially) satisfy the requirements for \revm{working and design phases}.} An analysis of promising contributions in \revm{the} light of the requirements pinpointed in Section~\ref{sec:challenges-subsec} is depicted in Table~\ref{tab:prospects}. \rev{Specifically, four families of techniques offer compelling research opportunities in this direction:}

\begin{mdframed}[style=prospects,frametitle={\rev{\textbf{Prospects}: Consensus}}]
	\rev{As illustrated in Section~\ref{sec:consensus}, in very simple settings (e.g., a system of dynamically decoupled \revM{subsystems} with single integrator dynamics), a consensus approach offers a solution that is scalable to \revm{an ULS}. Two directions are worth exploring in this context:
	\begin{itemize}
		\item A performance-driven design of consensus-based control solutions requires, in general, the solution of a Lyapunov equation, which is not scalable. However, research on scalable performance-driven design seems to have been largely overlooked in the literature. Recently, in \cite{Jiao2020,BarkaiMirkinEtAl2022,DijCha_CDC23a} this prospect has been identified and preliminary efforts have been carried out in this promising direction;
		\item The application of a consensus solution in more complex systems generally entails (see, e.g., \cite[Chap.~8]{RenBeard2007} for more details): (i)~\revm{the} identification of information, called the \emph{coordination variable}, that, if know in every \revM{subsystem}, would allow \revm{for} coordination (e.g., the common heading in a flocking problem); (ii)~\revm{the centralized design of} a distributed control law assuming that the coordination variable is know in every \revM{subsystem}; (iii)~\revm{the implementation of} a consensus algorithm to reach an agreement on the coordination variable, which is then used to implement the centrally designed control law. First, \revm{there is the opportunity to} \emph{\revm{develop} distributed design} \revm{procedures for the distributed control law of control problems that can be formulated in the aforementioned fashion, which}  would immediately suit the overall solution for \revm{an ULS}. Second, a performance-driven distributed design of the consensus protocol accounting for the distributed control law \revm{promises to} offer a significant performance improvement.
	\end{itemize}}
\end{mdframed}	

\begin{mdframed}[style=prospects,frametitle={\rev{\textbf{Prospects}: Cooperative Localization}}]
	\rev{On the one hand, simple cooperative localization algorithms such as \cite{Carrillo-ArceNerurkarGordilloEtAl2013,WanasingheEtAl2014} satisfy all of the \revm{ULS} feasibility requirements. However, that is achieved at the expense of very conservative estimates. On the other hand, more complex algorithms that aim at reducing conservativeness have been proposed, such as \cite{LuftEtAl2018}, but they do so at the expense of \revm{underestimating the correlation between measurements (double-counting)} and/or unscalable resource usage. There is a gap to develop well-performing cooperative localization algorithms explicitly accounting for the \revm{ULS} requirements outlined in Section~\ref{sec:challenges-subsec} that are both \emph{\revm{immune to double-counting} and scalable to \revm{an ULS}}.}
\end{mdframed}	

\begin{mdframed}[style=prospects,frametitle={\rev{\textbf{Prospects}: Clustering}}]
	\rev{As illustrated in Section~\ref{sec:clustering}, given a static clustering configuration, both the design and working phases are scalable to \revm{an ULS}. For increased performance, oftentimes \revm{a} hierarchical control approach is taken whereby dynamic clustering is performed on a high level layer (e.g., \citet{MaestreMunozdelaPenaEtAl2014,ChanfreutMaestreEtAl2021}). Nevertheless, state-of-the-art dynamic clustering algorithms have to be designed on a central entity. Thus, there is a clear gap to develop distributed dynamic clustering algorithms that are scalable to \revm{an ULS}.}
\end{mdframed}

\begin{mdframed}[style=prospects,frametitle={\rev{\textbf{Prospects}: Distributed Model Predictive Control}}]
	A promising approach seems to have been unveiled by research on DMPC schemes, namely \citet{RichardsHow2004,RichardsHow2004b, KeviczkyBorrelliEtAl2006, Dunbar2006, Dunbar2007, Zhao2015, PedrosoBatista2022}. Despite these \revm{existing} contributions, which focus on the particular case of \revM{subsystems} with decoupled dynamics or restrictive network topologies, great strides still have to be undergone to meet the \rev{design and scalability requirements for the} envisioned \revm{ULS} applications. \rev{First, there is a gap for scalable DMPC approaches suitable in more general settings, such as:
	\begin{itemize}\itemsep0em
		\item Dynamically coupled \revM{subsystems};
		\item Time-varying interaction typologies.
	\end{itemize}
	Second, there is a gap for scalable methods with guarantees such as:
	\begin{itemize}\itemsep0em
		\item Stability;
		\item Safety;
		\item Performance.
	\end{itemize}
	Very} few solutions that satisfy the \revm{ULS} requirements in Table~\ref{tab:req} have stability guarantees. On top of that, to the best of our knowledge, no solutions exist that satisfy the \revm{ULS} implementation feasibility requirements, even for a particular case with decoupled dynamics, that have safety or performance guarantees. \rev{The MPC approach has historically proven to be very versatile in enforcing \revm{safety and performance} guarantees \revm{\citep{HewingKabzanEtAl2020,BerberichKoehlerEtAl2021,DuanDrekeEtAl2024}.}}
\end{mdframed}		

\subsection{\rev{Consistency: The key for a distributed design}}

All the works featured in Table~\ref{tab:prospects} rely on a decomposition of a network-wise control problem into local problems in each \revM{subsystem}\rev{, which are} solved resorting to local communication. A critical step of such a decomposition is how each \revM{subsystem} captures the dependence of the local solution of the remainder of the network on its local \rev{design} procedure. \rev{In this section, we explore this concept, \revm{called consistency,} and we argue that it is key to obtain guarantees in the distributed design phase.}

In general, \revm{a real-time distributed design procedure in each subsystem uses a local approximation of the local contribution towards the network-wise objective, which is computed with access to partial information. A distributed \rev{design} procedure is said to be \emph{consistent} if, in each \revM{subsystem}, such approximation upper bounds the local ground-truth contribution.} \rev{We borrow the} \revm{the term \emph{consistency}} from the estimation literature \citep{Julier1997,julier2017general}. In the literature of DMPC, this property is often \revm{called} consistency or compatibility of local subproblems \citep{muller2017economic}.

\begin{mdframed}[style=example,frametitle={\rev{\textbf{Example}: Satellite constellation (running example)}}]
Consider, as an illustrative example, an estimation problem on the network of satellites of  Fig.~\ref{fig:eg_sat}, with the goal that each satellite estimates its own state, given relative position measurements between adjacent satellites, and  communication of local estimates among adjacent satellites. \revm{For inertial positioning, some of the satellites, commonly denominated as anchors, have access to their own inertial position.} In this case, the computation of a gain of, for instance, a local Luenberger filter in satellite $\mathcal{S}_i$ is highly dependent on the correlation between the state estimation errors of $\mathcal{S}_i$ and any adjacent satellite $\mathcal{S}_j$. In a centralized \rev{design} setting, such correlation is exactly known (assuming linear dynamics). Nevertheless, for a generic network topology, such correlation information cannot be kept if one is constrained by the \revm{ULS} feasibility requirements presented in Section~\ref{sec:challenges}, see \cite[Section~3.3]{PedrosoBatista2023b} for a detailed analysis. As a result, distributed \rev{design} procedures must capture such correlation approximately. Therefore, the knowledge of the \revm{covariance of the} state estimation error of $\mathcal{S}_i$ is also approximate. \rev{\revm{In this case, we} say} that the distributed \rev{design} procedure is \emph{consistent} if the ground-truth \revm{covariance of the} state estimation error in $\mathcal{S}_i$ is upper bounded by the \revm{one computed in $\mathcal{T}_i$}. An analogous \revm{analysis}, albeit not as obvious, can be carried out for the satellite constellation \revM{formation} control problem of the network in Fig.~\ref{fig:eg_sat}.
\end{mdframed}

\rev{Essentially, to achieve stable cooperation, the local design phase requires information about the impact that the interconnection with other \revM{subsystems} has on the network-wise objective.} \rev{Thus, it is not a surprise that sufficient conditions for stability often rely on a consistency concept (see, e.g., \citet[Section~4]{KeviczkyBorrelliEtAl2006}). Indeed, the only DMPC approaches that satisfy the \revm{ULS} implementation feasibility requirements and have stability guarantees are the ones that have a consistency property \citep{Dunbar2006,Dunbar2007}. Indeed,} neither the scheme in \cite{KeviczkyBorrelliEtAl2006} nor in \cite{PedrosoBatista2023b,PedrosoBatista2022}, \rev{which are not provably stable}, are guaranteed to be consistent. 

\begin{mdframed}[style=callout]
	\rev{\revm{The} notion of consistency can be enforced by locally formulating the design phase as a robust optimization problem that accounts for the worst-case impact of other \revM{subsystems} on the network-wise objective, given locally available information. This framework has been used in the estimation literature for two decades (e.g., \citet{ArambelRagoEtAl2001,LiNashashibi2013}), and may prove fruitful in the development of the next generation of scalable control solutions.}
\end{mdframed}

It is \rev{also} important to stress the meaningfulness of this property \rev{from a supervisory point of view}. Indeed, if a scheme is not consistent, it is not possible to locally assess the quality of the control or estimation solution in real time. For example, in the illustrative satellite estimation problem above, it is not possible to assess, in real time, the upper bound of the state estimation error for each satellite, which may have dire consequences (e.g., in debris avoidance maneuvers).

\section{Conclusions}\label{sec:conclusion}

The quest for resilience, flexibility, and scalability in urgent engineering applications is inducing a paradigm shift towards architectures of \revm{an ultra} large number of \revm{interconnected simpler subsystems}. However, beyond a certain scale, \revd{\emph{centralized}} \rev{design} methods for \rev{decentralized and distributed} controllers\revm{, which require some form of global information,} are no longer feasible. \revm{This calls for} new disruptive techniques for this emerging class of systems, which we \revd{refer to as} \revm{\emph{ultra large scale systems} (ULSS)}. Indeed, the local communication, computational, and memory requirements of the design and working phases of control algorithms deployed \revd{on} \revm{an ultra large scale (ULS)} are subject to strict constraints for real-time implementation feasibility. 

\revd{First}, \rev{\revm{ULS}} requirements and demand for suitable control algorithms are underpinned with examples of pressing \revm{ULS} large-scale applications.

\revm{\revd{Second}, the ULS requirements are formally identified. These requirements establish that the quantity of the local resources required in each subsystem to perform both working and design phases is bounded as the number of subsystems grows unbounded.}

\revm{\revd{Third}, state-of-the-art decentralized and distributed control approaches are reviewed in the light of the \revd{identified} ULS requirements. The following conclusions were drawn:
\begin{itemize}\itemsep0em
\item Consensus arguments provide a scalable methodological solution for control problems of ULSS of dynamically decoupled subsystems. Nevertheless, \revd{once optimality according to some performance criteria is added (instead of just consensus), the} design phase \revd{becomes} challenging to distribute among subsystems on an ULS;
\item $\mathcal{H}_2/\mathcal{H}_\infty$, convex relaxation, and special-structure systems design approaches often allow for a convex design problem, but are mostly not suitable to be distributed among the subsystems \revd{on} an ULS;
\item Clustering-based approaches follow the ULS requirements for a given \emph{precomputed} clustering configuration. If the clustering is performed dynamically, resorting to a multi-layer controller, state-of-the-art clustering algorithms are centralized and require some form of global information.
\end{itemize}}

\revm{Finally, from} the comparison between the identified technical requirements and the state-of-the-art approaches to this problem, it can be concluded that \revd{these requirements} have not yet been addressed adequately, presenting very compelling and potentially impactful future research directions. The most promising are: 
\begin{itemize}\itemsep0em
	\item \revm{Develop} algorithms that abide by all strict \revm{ULS} requirements for relevant settings \revd{that show potential but have not fully matured in the} in the literature, \revm{namely: performance-driven design of consensus protocols, dynamic clustering algorithms on a multi-layer control approach, and distributed model predictive control schemes;} 
	\item Establish \revm{guarantees (e.g., stability, safety, and performance) for control approaches that are scalable on an ULS}. \revm{Establishing consistency of local} approximations employed to distribute the \rev{design} load among the \revM{subsystems} \revm{seems to be a promising principle}.
\end{itemize}

\revd{ULSS offer a highly relevant research domain for control science and engineering that will play important roles in many societal domains in the future. New control design methodologies have to be conceived and, through this vision paper, we intended to spark a further interest of the control community in this relevant research area.}


	\section*{Acknowledgments}
	The authors thank Dr.\ Ilse New for proofreading this paper \revm{and two anonymous reviewers for numerous comments that allowed for substantial improvements.} This work was partially supported by LARSyS \revm{FCT} funding (DOI: {\small \href{https://doi.org/10.54499/LA/P/0083/2020}{\texttt{10.54499/LA/P/0083/2020}}, \href{https://doi.org/10.54499/UIDP/50009/2020}{\texttt{10.54499/UIDP/ 50009/2020}}}, and {\small \href{https://doi.org/10.54499/UIDB/50009/2020}{\texttt{10.54499/UIDB/50009/2020}}}).
\appendix

\bibliographystyle{elsarticle-harv} 
\bibliography{../../_bib/references-c.bib}

\end{document}